\documentclass[journal]{IEEEtran}
\usepackage{cite}
\usepackage{amsmath,amssymb,amsfonts}
\usepackage{bm}
\usepackage{algorithmic}
\usepackage{graphicx}
\usepackage{subfigure}
\graphicspath{{figures/}}
\usepackage{textcomp}
\usepackage[ruled,linesnumbered]{algorithm2e}
\usepackage{booktabs}
\usepackage{multirow}

\usepackage{mathtools}

\usepackage{hyperref}
\usepackage{wasysym}
\usepackage{utfsym}
\usepackage{amsthm} 
\usepackage{cases}
\usepackage{tabularx}

\usepackage{wasysym}
\usepackage{utfsym}

\begin{document}
\let\bs\boldsymbol
\newcommand{\incmk}{\in\mathbb{C}^{M_k}}
\newcommand{\inc}{\in\mathbb{C}}

\newcommand{\N}{N}
\newcommand{\K}{N}
\newcommand{\mK}{\mathcal{K}}

\newcommand{\mNk}{\mathcal{N}_k}
\newcommand{\mNl}{\mathcal{N}_l}

\newcommand{\Nk}{N_k}
\newcommand{\Nl}{N_l}
\newcommand{\Mk}{M_k}
\newcommand{\Qk}{Q_k}

\newcommand{\Pnk}{P_{n_k}}

\newcommand{\nk}{n_k}

\newcommand{\MSE}[1]{MSE_{x,n}\left\{{#1}\right\}}

\newcommand{\hnlk}{\bs{h}_{n_l,k}}
\newcommand{\hnkk}{\bs{h}_{n_k,k}}

\newcommand{\unk}{u_{n_k}}
\newcommand{\unl}{u_{n_l}}

\newcommand{\ank}{a_{n_k}}
\newcommand{\anl}{a_{n_l}}
\newcommand{\tnk}{\theta_{n_k}}
\newcommand{\xni}{x_{n_i}}
\newcommand{\xnk}{x_{n_k}}
\newcommand{\xnl}{x_{n_l}}

\newcommand{\n}{\bs{n}}

\newcommand{\z}{\bs{z}}  

\newcommand{\yk}{{\bs{y}}_k}  

\newcommand{\xhk}{\hat{x}_k}
\newcommand{\xk}{x_k}
\newcommand{\vk}{\bs{v}_k}
\newcommand{\vkl}{\bs{v}_k^{(l)}}

\newcommand{\dnik}{d_{{n_i},k}}

\newcommand{\unks}{{\unk^*}}

\newcommand{\vko}{{\vk^{\text{opt}}}}
\newcommand{\unio}{{\uni^{\text{opt}}}}
\newcommand{\unko}{{\unk^{\text{opt}}}}
\newcommand{\tks}{{\tk^*}}
\newcommand{\abs}[1]{\lvert{{#1}}\rvert}

\newcommand{\vks}{{\vk^*}}
\newcommand{\tk}{t_k}
\newcommand{\tko}{{t^o_k}}

\newcommand{\norm}[1]{{\lVert{#1}\rVert}}

\title{Unfolded Deep Graph Learning for Networked Over-the-Air Computation}

\author{Xiao Tang, Huirong Xiao, Chao Shen, Li Sun, Qinghe Du, Dusit Niyato, and Zhu Han%
\thanks{X. Tang is with the School of Information and Communication Engineering, Xi'an Jiaotong University, Xi'an 710049, China, and also with Research \& Development Institute of Northwestern Polytechnical University in Shenzhen, Shenzhen 518063, China. (e-mail: tangxiao@xjtu.edu.cn)}
\thanks{H. Xiao is with the School of Electronics and Information, Northwestern Polytechinical University, Xi'an 710072, China.}
\thanks{C. Shen is with the Faculty of Electronic and Information Engineering, Xi'an Jiaotong University, Xi'an 710049, China.}
\thanks{L. Sun and Q. Du are with the School of Information and Communication Engineering, Xi'an Jiaotong University, Xi'an 710049, China.}
\thanks{D. Niyato with the School of Computer Science and Engineering, Nanyang Technological University, Singapore.}
\thanks{Z. Han is with the Department of Electrical and Computer Engineering, University of Houston, Houston 77004, USA.}
}

\maketitle

\begin{abstract}
Over-the-air computation (AirComp) has emerged as a promising technology that enables simultaneous transmission and computation through wireless channels. In this paper, we investigate the networked AirComp in multiple clusters allowing diversified data computation, which is yet challenged by the transceiver coordination and interference management therein. Particularly, we aim to maximize the multi-cluster weighted-sum AirComp rate, where the transmission scalar as well as receive beamforming are jointly investigated while addressing the interference issue. From an optimization perspective, we decompose the formulated problem and adopt the alternating optimization technique with an iterative process to approximate the solution. Then, we reinterpret the iterations through the principle of algorithm unfolding, where the channel condition and mutual interference in the AirComp network constitute an underlying graph. Accordingly, the proposed unfolding architecture learns the weights parameterized by graph neural networks, which is trained through stochastic gradient descent approach. Simulation results show that our proposals outperform the conventional schemes, and the proposed unfolded graph learning substantially alleviates the interference and achieves superior computation performance, with strong and efficient adaptation to the dynamic and scalable networks.
\end{abstract}


\section{Introduction}
With the evolution towards the sixth generation (6G) wireless communications, the wireless networks are envisioned to accommodate unprecedented number of devices with numerous applications. As such, the wireless networks are expected for not only efficient data transmission but also intelligent data processing to support various services~\cite{6g}. Towards this vision, the conventional transmit-and-process protocol is severely challenged by the vast number of devices and the huge amount of data due to the immense spectrum and power resources exhausted therein~\cite{dang1}. Recently, the over-the-air computation (AirComp) has emerged as a promising technology to enable efficient data aggregation and processing directly through the wireless medium~\cite{comp-cpt}. Particularly, AirComp leverages the superposition property of multiple-access channels to perform computation on the wireless signals, and thus enables simultaneous transmission and processing of data~\cite{w3}. As computation is conducted over the air, data processing is facilitated with improved efficiency and scalability while conserving the resources, allowing the wireless operators with more prosperous functionalities rather than simply laboring as the data pipelines~\cite{comp-princ}.

Due to the unique way of AirComp in terms of data processing, it finds significant potentials in various scenarios, such as sensing data fusion, control information aggregation, and federated learning, covering a wide variety of cyber-physical system applications~\cite{aircomp}. As such, it is critical to extend the existing single-task AirComp to the networked AirComp allowing concurrent computation instances of different services. From a networked perspective, it can be modeled under a multi-cluster AirComp architecture, where each cluster is associated with an independent fusion center~\cite{aircomp-mult}. In this regard, different types of tasks such as temperature measurement, fire monitoring, and neural network learning, can be conducted concurrently~\cite{aircomp-netw}. However, the coexistence of multiple clusters introduces the interference, which poses a significant challenge as it affects the data aggregation in each cluster and further hinders the scalability to adopt more diversified data. Therefore, besides the transceiver coordination issue in conventional AirComp scenarios, the interference needs to be managed properly towards effective networked AirComp~\cite{aircomp-interf}.

Moreover, the principle of AirComp requires joint optimization at the transmitter and receiver sides to achieve effective computation. Accordingly, efficient and accurate determination of transceiving strategy becomes the fundamental support~\cite{ac-opt}. However, when addressing this issue from the optimization perspective, it usually requires multi-round iterations to achieve some stationary-point strategies, which may not necessarily be the optimal and is also time-consuming. Meanwhile, machine learning techniques features prompt execution through a well-trained neural network, yet the pure data-driven approaches may not have sufficient adaptability in changing environments~\cite{w1}. In this context, the unfolded learning technique featuring joint knowledge- and data-driven designs emerges as an attractive solution~\cite{alg-uf}. Guided by the derivations from theoretical perspectives, the unfolding technique leads to a neural network architecture inspired by problem-oriented iterations. Accordingly, the unfolded learning can effectively interpret the problem structure through promptly executable neural networks with improved adaptivity and scalability~\cite{uf-eg}.

Meanwhile, for the networked AirComp with multi-cluster concurrent computation, the strategy determination of one cluster influences those of the rest~\cite{ac-inter}. From a learning perspective, the graph neural network (GNN) naturally fitting graph-structure data is of particular interest to tackle problems in such forms~\cite{gnn-concp}. Through a GNN framework, the transceiving behavior in networked AirComp can be embedded as the features of graph elements, and the mutual influence among different clusters can be implanted within the graph structure. Moreover, the desired feature of permutation equivalence in GNNs allows convenient generalization to arbitrary scale of networks~\cite{gnn-adv}. Therefore, taking advantage of GNN within an unfolded learning architecture provides an effective tool to tackle the networked AirComp issue, where a well-trained neural network can be capable of providing effective computation strategies, while scaling to dynamic network environments~\cite{gnn-uf}.

Motivated by the vision towards networked AirComp allowing diverse and concurrent computation, in this paper, we investigate the multi-cluster AirComp by unfolded deep graph learning. Inspired by the optimization-based decomposition and analysis, we propose an unfolded learning framework exploiting the GNN structure for efficient and generalizable computation strategy design. Specifically, the main contribution can be summarized as follows:
\begin{itemize}
	\item We investigate a multi-cluster AirComp network where each cluster is allowed with its own AirComp tasks. The coexistence of clusters incurs mutual interference where we aim to maximize the weighted-sum AirComp rate by addressing the transceiver design.
	\item From an optimization perspective, we decompose the weighted-sum AirComp rate maximization into transmission and reception subproblems. Through an alternating optimization framework, the subproblems are solved with successive convex approximation, and iterated to approach the optimal networked AirComp strategy.
	\item Induced by the theoretical analysis, we propose an unfolded learning architecture where the iterative optimization procedure is reinterpreted with cascaded neural network-based solving blocks. {We adopt a} GNN with message passing mechanism in each block to track the interactive transceiving behavior among clusters with mutual interference.
	\item The simulation indicates that the unfolded learning scheme effectively controls the network interference and thus outperforms the conventional schemes with more efficient execution. Moreover, the trained neural network can be conveniently generalized to various AirComp network scenarios with guaranteed performance.
\end{itemize}

The rest of this paper is organized as follows. In Sec.~\ref{sec:rw}, we review the related work. Sec.~\ref{sec:sys} introduces the networked AirComp system, and Sec.~\ref{sec:opt} analyzes the considered AirComp problem through optimization techniques. In Sec.~\ref{sec:nn}, the problem-specific unfolded graph learning is proposed based on the theoretical analysis presented before. Sec.~\ref{sec:sim} provides the simulation results, and finally Sec.~\ref{sec:con} concludes this paper.

\section{Related Work} \label{sec:rw}

As a promising technique towards resource-efficient integration of communication and computation, AirComp has attracted increasing research attention recently~\cite{aircomp}. As an alternative way for wireless data fusion, AirComp has been investigated under various scenarios, such as high-mobility multi-modal sensing~\cite{sens}, integrated sensing and communications~\cite{isac}, reconfigurable intelligent surface-assisted transmissions~\cite{ris}, wireless-powered communication~\cite{wpt}, amplify-and-relay transmissions~\cite{relay}, and unmanned aerial vehicle networks~\cite{uav}. In these studies, the coordinated transmissions and receptions are jointly investigated along with the particular scenario-related settings towards the optimal computation strategy. However, when addressing AirComp in a networked scope, the interference management becomes a critical issue~\cite{aircomp-interf}. In~\cite{ac-inter}, the authors introduce the interference temperature constraint over multi-cell AirComp to reach distributed interference management for Pareto-optimal computation policy. In~\cite{inter-1}, the authors address the inter-cluster interference and non-uniform fading, and propose a uniform-forcing transmitter design for decentralized and low-complexity AirComp solutions. In~\cite{inter-2}, the authors investigate the fair AirComp over multiple spectrum-sharing clusters and reveal that inter-cell interference diminishes with increased receive antennas. In~\cite{inter-3}, the authors introduce a MapReduce-based AirComp of large-scale nomographic functions, improving computation efficiency by harnessing the interference rather than combating it. 

While the aforementioned work provides effective design for AirComp in various cases, they basically exploit analytical techniques which usually requires some time-consuming iterative processes to find the solution. In this regard, the machine learning-based approaches provide an intelligent alternative for AirComp designs~\cite{ac-nn}. In~\cite{enc}, the authors present an end-to-end communication system for AirComp, leveraging autoencoder network structures to for transceiver design. In~\cite{edge}, the authors investigate the task-oriented AirComp for multi-device edge intelligence systems, specifically focusing on improving inference accuracy in split-inference systems. In~\cite{know}, the authors address the AirComp-assisted federated learning, introducing a knowledge-guided approach to improve the convergence and accuracy. In~\cite{airnn}, the authors leverage reconfigurable intelligent surface-assisted AirComp for convolutional neural network computation in the analog domain. In~\cite{drl}, the authors propose a distributed deep reinforcement learning framework for optimizing unmanned aerial vehicle swarm-assisted AirComp for transportation systems.

Deep unfolding capitalizing on principled algorithm for model-driven learning has emerged as an attractive approach to make the neural network better fit the problem-specific structure for more effective learning~\cite{alg-uf}. Accordingly, the unfolding technique has been applied in multi-antenna precoding~\cite{uf-prec}, signal detection~\cite{uf-detec}, reconfigurable intelligent surface-aided communications~\cite{uf-ris}, unmanned aerial vehicle communications~\cite{uf-uav}, etc., towards knowledge and data dual driven schemes. When addressing the problems in a networked scope, graph appears as a powerful tool to track the mutual interactions for unfolding algorithm design~\cite{gnn-concp}. In~\cite{uf-gnn-wmmse}, the authors proposed to unfold the classical weighted minimum mean square error (WMMSE)-based method while integrating GNNs for rate maximization. The authors in~\cite{uf-gnn-coord} {extended} the previous approach to multi-cell multi-user scenarios to achieve coordinated network transmissions. In~\cite{uf-gnn-pwr}, the authors {proposed} the unfolded WMMSE with attentive graph representation to optimize power distribution in ad hoc wireless networks. In~\cite{uf-gnn-ee}, the authors introduced a graph-based trainable framework to unfold the successive concave approximation to maximize the weighted-sum energy efficiency.

\section{System Model} \label{sec:sys}


\begin{figure}
	\centering
	\includegraphics[width=\linewidth]{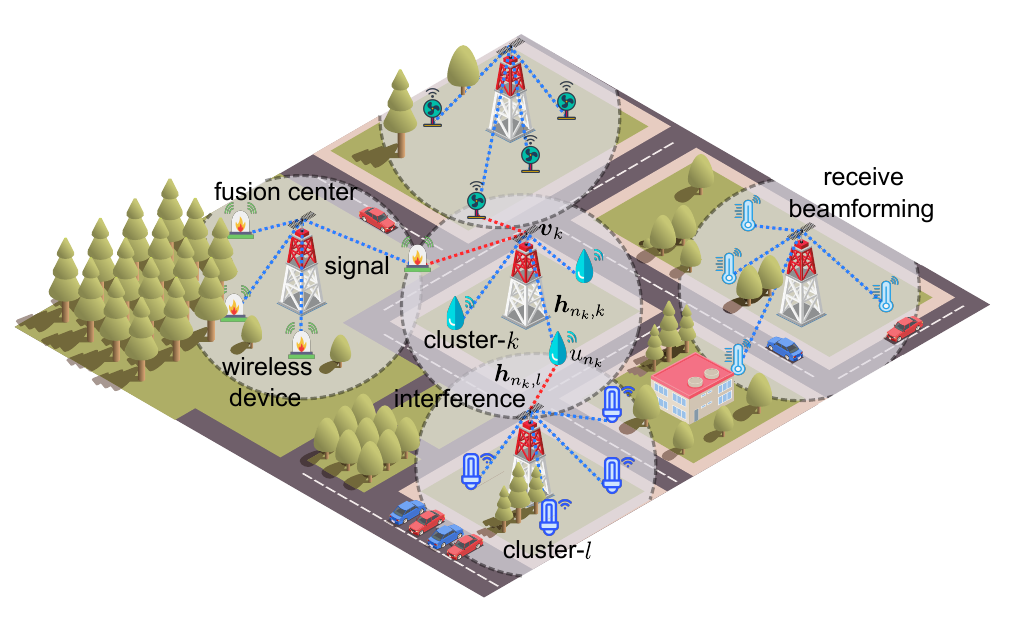}
	\caption{Networked AirComp system.}
	\label{fig:system-model}
\end{figure}

As illustrated in Fig.~\ref{fig:system-model}, we consider an AirComp network that comprises $K$ clusters, each cluster is of its own intended AirComp task, such as temperature, humidity, traffic monitoring, where the set of all clusters is denoted by $\mathcal{K}=\{1,2,\hdots,K\}$. The multi-cluster structure is motivated by the need to support diverse and concurrent computation tasks. In each cluster, there is a fusion center which aggregates the encoded wireless signals from the sensorial devices within the cluster for information inference. Also, there consists of $N_k$ wireless devices in the cluster-$k$, denoted by $\mNk=\{1,{\ldots},n_k,{\ldots},\Nk\}$ such that $n_k$ indicates the $n$-th device in the {cluster-$k$}. In this work, we assume that the association of devices to the fusion center is pre-determined, which can be conveniently implemented based on computation task or geographic proximity. In the AirComp network, the sensorial devices are equipped with one single antenna for cost efficiency, while the fusion centers have multiple antennas for more effective signal processing, where the number of antennas at the fusion center in cluster-$k$ is denoted by $M_k$. Then, for the AirComp signal transmissions, the channel condition from device-$n_l$ in cluster-$l$ to the fusion center of cluster-$k$ is specified as $\hnlk \incmk$, which corresponds to the intra-cluster communication when $ l $ equals $k$, and otherwise, inter-cluster interference occurs. Meanwhile, the transmission scalar from device-$n_k$ in cluster-$k$ is denoted by $\unk$, and is subject to the power constraint that
\begin{equation}
	\left\lvert\unk \right\rvert^2 \leq \Pnk, \quad \forall n_k\in\mathcal{N}_k,\quad \forall k\in\mathcal{K},
	\label{eq:power_constrain}
\end{equation}
where $\Pnk $ is the maximum transmit power of device-$n_k$. Also, the receive beamforming at the fusion center in cluster-$k$ is represented by $\vk\incmk$.

For the considered AirComp network, all the clusters with their devices conduct AirComp tasks concurrently. As such, the received signal at the fusion center in cluster-$k$ is
\begin{equation} \label{eq:yk}
	\yk=\sum\limits_{l\in\mathcal{K}}\sum\limits_{n_l\in\mathcal{N}_l}\hnlk \unl \xnl + \z_k, \quad \forall k\in\mathcal{K},
\end{equation}
where $ \xnl $ denotes the transmitted symbol from device-$n_l$ in cluster-$l$ with $ \xnl\sim\mathcal{CN}(0,1) $, and $ \z_k $ is the noise at the fusion center in cluster-$k$ with $ \z_k \sim \mathcal{CN}(\bm{0}, \sigma_k^2\bm{I}_{M_k}) $. We can see that the transmissions from all clusters are overlapped and thus intended aggregation in each cluster is affected by the inter-cluster interference. Through the received superposed signal, the fusion center in each cluster conducts the computation of a certain type of nomographic function that fits its intended task~\cite{Nomographic}. Here we assume perfect synchronization and channel state information is available to support the networked AirComp operations. This can be achieved through standard procedure like preamble-based training and thus is a commonly adopted in existing work~\cite{aircomp,aircomp-interf,lee1,li2}. In this paper, we adopt the sum operation as the interested nomographic function, which can also be conveniently extended to the cases with functions of other types. Particularly, as each fusion center is only interested in AirComp over the devices of its own service, the coexisted transmissions from other clusters are considered as interference. For the fusion center in cluster-$k$ with sum operation, it expects to obtain
\begin{equation} \label{eq:xk}
	s_k = \sum\limits_{n_k\in\mathcal{N}_k}\xnk, \quad \forall k\in\mathcal{K},
\end{equation}
from the received signal in~(\ref{eq:yk}). To this end, the fusion center conducts receive beamforming over the aggregated signal to facilitate the recovery of its intended signal. Then, the actually obtained aggregation at fusion center-$k$ becomes
\begin{equation} \label{eq:xhk}
	\hat{s}_k = \vk^H\yk, \quad \forall k\in\mathcal{K},
\end{equation}
with $ \vk $ being the receive beamformer therein.

Given the evident difference between the expected and actually aggregated results in~(\ref{eq:xk}) and~(\ref{eq:xhk}), respectively, we adopt the mean square error (MSE) of the difference as the performance metric, given as
\begin{equation} \label{eq:msek}
	\mathsf{MSE}_k = \mathbb{E}\left\{\left\lvert s_k-\hat{s}_k\right\rvert^2\right\}, \quad \forall k\in\mathcal{K},
\end{equation}
for cluster-$k$. Then, by substituting the expression of the received signals, the MSE in~(\ref{eq:msek}) is extended as
\begin{equation} \label{eq:msek_ex}
\begin{aligned}
	\mathsf{MSE}_k = & \sum\limits_{n_k\in\mathcal{N}_k} \left\lvert\vk^H \hnkk \unk-1\right\rvert^2 \\
		& + \sum\limits_{l\in\mathcal{K}}\sum\limits_{n_l\in\mathcal{N}_l} \left\lvert\vk^H \hnlk \unl \right\rvert^2
		+\vk^H\vk\sigma_k^2, \:\: \forall k\in\mathcal{K},
\end{aligned}
\end{equation}
where the first term originates from the intra-cluster transmission, the second and last terms correspond to the side effect due to interference and noise, respectively. Evidently, the smaller MSE becomes, the closer the actual computation approaches the target. Note the analysis above is built upon the sum operation during AirComp, while it can be conveniently extended to the cases with other nomographic computation with the MSE metric derived in similar manners. Despite the effectiveness of MSE to evaluate the AirComp performance, it remains a cluster-wise metric and can hardly be extended to a networked scope. In this respect, we adopt the recently introduced concept of AirComp rate, which is capable of the scenarios with concurrent AirComp operations~\cite{comp-princ,comp-rate}. Particularly, with achieved MSE in~(\ref{eq:msek}) for cluster-$k$, the AirComp rate corresponding to the number of computed function values per channel use in num/Hz is then obtained as
\begin{equation}
	r_k = \frac{1}{Q_k+\log_2\Nk}\log_2^+\left(\frac{1}{\mathsf{MSE}_k}\right), \quad \forall k\in\mathcal{K},
	\label{eq:rate}
\end{equation}
where $Q_k$ is the number of the optimal quantization bits for the type of computation function and $N_k$ is the number of wireless devices in this cluster, and $\log^+_2(\cdot) \triangleq \max\{\log_2(\cdot), 0\}$. Given the monotonic relationship between the MSE and the AirComp rate, we can see that a smaller MSE induces a higher AirComp rate. Therefore, for the concurrent AirComp in multiple clusters, we intend to maximize the weighted-sum AirComp rate, specified as
\begin{IEEEeqnarray}{cl}\label{eq:raw_ploblem}
	\IEEEyesnumber \IEEEyessubnumber*
	\max_{\substack{\{\vk\}_{k\in\mathcal{K}} \\ \{u_{n_{{l}}}\}_{n_{{l}}\in\mathcal{N}_{{l}}, {l}\in\mathcal{K}}}}
	&\quad \sum\limits_{k\in\mathcal{K}} w_kr_k
	\label{eq:17a}\\
	\mathrm{s.t.}
	&\quad\left\lvert u_{n_{{l}}} \right\vert^2 \leq P_{n_{{l}}}, \quad \forall n_{{l}}\in\mathcal{N}_{{l}}, \:\: \forall {l}\in\mathcal{K},
	\label{eq:constrain} \IEEEeqnarraynumspace
\end{IEEEeqnarray}
where $ \{w_k\}_{k\in\mathcal{K}} $ are the associated weights for different clusters. Through the formulated problem, we then jointly investigate the transceiver design in all clusters to improve the networked AirComp.

\section{An Optimization Perspective} \label{sec:opt}

For the formulated problem in~(\ref{eq:raw_ploblem}) to tackle networked AirComp, we first analyze it from an optimization perspective. Accordingly, we need to jointly optimize the transmission scalar from all sensorial devices and the receive beamforming at all fusion centers in the network. In this regard, the mutual interference coupling among different clusters induces a major difficulty towards efficient solutions. Towards this issue, we then discuss the problem decomposition and propose the algorithm design based on the alternating optimization technique.

For the networked AirComp problem in~(\ref{eq:raw_ploblem}), we can readily see that the transmit and receive design can be tackled separately. Particularly, the subproblem for receive beamforming with given transmit scalars is specified as
\begin{equation}
	\max_{\{\vk\}_{k\in\mathcal{K} }} \quad \sum\limits_{k\in\mathcal{K}} w_k\frac{1}{Q_k+\log_2\Nk}\log_2^+\left(\frac{1}{\mathsf{MSE}_k}\right),
	\label{eq:sp1}
\end{equation}
while the transmit scalar optimization with fixed receive beamformer is given as
\begin{IEEEeqnarray}{cl} \label{eq:sp02}
	\IEEEyesnumber \IEEEyessubnumber*
	\max_{\{u_{n_{{l}}} \}_{n_{{l}}\in\mathcal{N}_{{l}}, {l}\in\mathcal{K}}} 	
	& \sum\limits_{k\in\mathcal{K}} w_k\frac{1}{Q_k+\log_2\Nk}\log_2^+\left(\frac{1}{\mathsf{MSE}_k}\right)	\\
	\mathrm{s.t.} 
	&\left\lvert u_{n_{{l}}} \right\rvert^2 \leq P_{n_{{l}}}, \quad \forall n_{{l}}\in\mathcal{N}_{{l}}, \forall {l}\in\mathcal{K}. \IEEEeqnarraynumspace
\end{IEEEeqnarray}
Then, for the unconstrained optimization in~(\ref{eq:sp1}), we can readily see that it can be solved in a cluster-wise manner, which corresponds to the MSE minimization in each individual cluster, given as
\begin{equation}
	\min_{\vk} \quad \mathsf{MSE}_k, \quad \forall k\in\mathcal{K}.
\end{equation}
Revisiting the obtained MSE in~(\ref{eq:msek_ex}) and rearranging the terms, we derive that
\begin{equation}
	\begin{aligned}
		\mathsf{MSE}_k
		=& \sum\limits_{l\in\mathcal{K}}\sum\limits_{n_l\in\mathcal{N}_l} \left\lvert\vk^H \hnlk \unl \right\rvert^2 + \vk^H\vk\sigma_k^2\\
		&-  \sum\limits_{n_k\in\mathcal{N}_k} (\vk^H\hnkk\unk+u^*_{n_k,k}\hnkk^H\vk) + N_k\\
		=\:&\vk^H \left[ \sum\limits_{l\in\mathcal{K}}\sum\limits_{n_l\in\mathcal{N}_l} \left\lvert\unl\right\rvert^2\hnlk\hnlk^H + \sigma_k^2\bs{I}_{M_k}\right]\vk\\
		&- 2\mathfrak{Re}\left(\vk^H \sum\limits_{n_k\in\mathcal{N}_k} \unk\hnkk\right) + N_k, \quad \forall k\in\mathcal{K},\\
	\end{aligned}
	\label{eq:mse_2}
\end{equation}
indicating the quadratic relation between the achieved MSE and receive beamforming. We can obviously see that in~(\ref{eq:mse_2}), $ \sum\nolimits_{l\in\mathcal{K}}\sum\nolimits_{n_l\in\mathcal{N}_l}\left\lvert\vk^H \hnlk \unl \right\rvert^2 + \vk^H\vk\sigma_k^2 > 0 $ is always satisfied regardless of $\vk$, and thus $ \sum\nolimits_{l\in\mathcal{K}}\sum\nolimits_{n_l\in\mathcal{N}_l} \left\lvert\unl\right\rvert^2\hnlk\hnlk^H + \sigma_k^2\bs{I}_{M_k} $ is a positive definite Hermitian matrix and thus invertible. In this regard, the MSE in~(\ref{eq:mse_2}) is convex with respect to the receive beamforming, and thus the optimal beamformer is obtained as
\begin{equation}
	\begin{aligned}
		\vk=&\left[ \sum\limits_{l\in\mathcal{K}}\sum\limits_{n_l\in\mathcal{N}_l} \left\lvert\unl\right\rvert^2\hnlk\hnlk^H + \sigma_k^2\bs{I}_{M_k}\right]^{-1}\\
		&\times\left[\sum\limits_{n_k\in\mathcal{N}_k}\unk\hnkk\right],\quad \forall k\in\mathcal{K}.	
	\end{aligned}
	\label{eq:optimal_vk}
\end{equation}

For the transmit scalar optimization subproblem in~(\ref{eq:sp02}), the inter-cluster interference induces mutually affected transmissions and thus the transmission from different clusters and devices need to be jointly addressed. Specifically, due to the non-concavity of the objective function, we introduce a set of slack variables $ \{t_k\}_{k\in\mathcal{K}} \ge 0 $ and reformulate~(\ref{eq:sp02}) as
\begin{IEEEeqnarray}{cl} \label{eq:subproblem2_sca}
	\IEEEyesnumber \IEEEyessubnumber*
	\max_{\substack{\{\tk\}_{k\in\mathcal{K} } \\ \{ u_{n_{{l}}}\}_{n_{{l}}\in\mathcal{N}_{{l}}, {l}\in\mathcal{K} }}}
	&\quad \sum\limits_{k\in\mathcal{K}} w_k\frac{1}{Q_k+\log_2\Nk}\log_2\left(\tk\right)\\
	\mathrm{s.t.}
	&\quad \left\lvert u_{n_{{l}}} \right\rvert^2 \leq P_{n_{{l}}}, \quad \forall n_{{l}}\in\mathcal{N}_{{l}}, \forall {l}\in\mathcal{K}, \label{eq:topt_con1}\\
	&\quad t_k>0, \quad \forall k\in \mathcal{K}, \label{eq:topt_con2}\\
	&\quad t_k \leq \frac{1}{\mathsf{MSE}_k}, \quad \forall k\in \mathcal{K}, \label{eq:topt_con3}
	\IEEEeqnarraynumspace\IEEEeqnarraynumspace\IEEEeqnarraynumspace
\end{IEEEeqnarray}
where the non-differential $ (\cdot)^+ $ operation is neglected. This generally leads to a lower-bounded AirComp rate as compared with the original, yet the approximation can be tight since the cases with negative AirComp rate can be safely ignored. Now the non-convexity lies in constraint~(\ref{eq:topt_con3}), which can be reorganized as $\mathsf{MSE}_k \leq {1}/{t_k}$, $ \forall k\in \mathcal{K} $. Then, by adopting successive convex approximation for the right-hand side at $\{\tk^\circ\}_{k\in\mathcal{K}}$, we have
\begin{equation}
	\frac{1}{t_k}\geq-\frac{1}{{(\tk^\circ)}^2}(\tk-\tk^\circ )+\frac{1}{\tk^\circ}=\frac{2}{\tk^\circ}-\frac{1}{{(\tk^\circ)}^2}\tk, \quad \forall k\in \mathcal{K}.
	\label{eq:relax_tk}
\end{equation}
For the MSE-related constraint, we can reorganize MSE against the interested optimization variable within a quadratic form and recast the constraint in~(\ref{eq:topt_con3}) as
\begin{multline}
	\left\lVert\begin{aligned}
		&\left[\vk^H\hnkk \unk-1\right]_{\nk\in \mathcal{N}_k}\\
		&\left[\vk^H\hnlk\unl\right]_{n_l\in \mathcal{N}_l, l\in \mathcal{K}\backslash\{k\} }\\
		&\frac{\frac{2}{\tk^\circ}-\frac{1}{{(\tk^\circ)}^2}\tk-\vk^H\vk\sigma_k^2-1}{2}
	\end{aligned}
	\right\rVert \\ \leq\frac{\frac{2}{\tk^\circ}-\frac{1}{{(\tk^\circ)}^2}\tk-\vk^H\vk\sigma_k^2+1}{2}, \quad \forall k\in \mathcal{K}.
	\label{eq:topt_con3_sca}
\end{multline}
which now becomes jointly convex with respect to $ \{\tk\}_{k\in\mathcal{K} } $ and $ \{\unl\}_{n_{{l}}\in\mathcal{N}_{{l}}, {l}\in\mathcal{K} } $. Therefore, we arrive at the convex counterpart of the transmit scalar optimization as
\begin{IEEEeqnarray}{cl} \label{eq:subproblem2_appx}
	\IEEEyesnumber \IEEEnosubnumber*
	\max_{\substack{\{\tk\}_{k\in\mathcal{K} } \\ \{u_{n_{{l}}}\}_{n_{{l}}\in\mathcal{N}_{{l}}, {l}\in\mathcal{K} }}}
	&\quad \sum\limits_{k\in\mathcal{K}} w_k\frac{1}{Q_k+\log_2\Nk}\log_2\left(\tk\right),\\
	\mathrm{s.t.}
	&\quad (\ref{eq:topt_con1}), (\ref{eq:topt_con2}), (\ref{eq:topt_con3_sca}), \nonumber
	\IEEEeqnarraynumspace\IEEEeqnarraynumspace\IEEEeqnarraynumspace
\end{IEEEeqnarray}
as approximated at $\{\tk^\circ\}_{k\in\mathcal{K}}$. We can then solve a series of convex problems in the form of~(\ref{eq:subproblem2_appx}) with updated approximated points with the obtained optimum, and the convergence leads to the solution to the transmit scalar problem in~(\ref{eq:sp02}).

As the transmit and receive designs are obtained through the decomposed subproblems, we can then adopt the alternating optimization technique to update the transceiver designs in an iterative manner to obtain the solutions for networked AirComp. The overall algorithm is summarized in Alg.~\ref{algorithm}, where the weight-sum AirComp rate is denoted by $ R = \sum\nolimits_{k\in\mathcal{K}}w_kr_k $ for notation simplicity and $ \epsilon_0 $ is the threshold claiming the convergence.

\SetKwRepeat{Do}{repeat}{until}
\begin{algorithm}[t] 
	\caption{Networked AirComp Optimization}
	\label{algorithm}
	\KwIn{ Network scenarios, $\{\hnlk\}_{n_l\in\mathcal{N}_l, k,l\in \mathcal{K} }$}
	\KwOut{$\{\vk\}_{k\in \mathcal{K}},\{\unl\}_{n_{{l}}\in\mathcal{N}_{{l}},{l}\in \mathcal{K} }$}
	Initialization: 
	randomly generate $\{u_{n_{{l}}}^0\}_{n_{{l}}\in\mathcal{N}_{{l}},{l}\in \mathcal{K}}$ and $\{\vk^0\}_{k\in\mathcal{K}}$ while satisfying the constraints\;
	\Do(\tcp*[f]{Alternating optimization}){$ \left| R^0 - R^\star \right| < \epsilon_0 $}{
		Calculate current weighted-sum {AirComp} rate as $ R_0 = R(\{u_{n_{{l}}}^0\}_{n_{{l}}\in\mathcal{N}_{{l}},{l}\in \mathcal{K}},\{\vk^0\}_{k\in\mathcal{K}}) $\;
		\For(\tcp*[f]{Receive beamforming}){$k\in\mK$}{
			Substitute $\{u_{n_{{l}}}^0\}_{n_{{l}}\in\mathcal{N}_{{l}},{l}\in \mathcal{K}}$ into~(\ref{eq:optimal_vk}), obtain receive beamformer for fusion center-$k$ as $\vk^\star$\;
			$ \vk^0 \gets \vk^\star $\;
		}
		{Calculate $\mathsf{MSE}_k$ according~(\ref{eq:mse_2}), let $t_k^o=\frac{1}{\mathsf{MSE}_k}$ for all $k$\;}
		\Do(\tcp*[f]{Transmit optimization}){$ \left| R^\circ - R^\star \right| < \epsilon_0 $} {
			Calculate current weighted-sum AirComp rate as $ R^\circ = R(\{u_{n_{{l}}}^0\}_{n_{{l}}\in\mathcal{N}_{{l}},{l}\in \mathcal{K}},\{\vk^0\}_{k\in\mathcal{K}}) $\;
			Formulate the problem in the form of~(\ref{eq:subproblem2_appx}), solve the problem to get $\{\tk^\star\}_{k\in\mathcal{K}}$ and $ \{u_{n_{{l}}}^\star\}_{n_{{l}}\in\mathcal{N}_{{l}},{l}\in \mathcal{K}} $, with the objective function updated as $ R^\star $\;
			$\tk^\circ \gets \tk^\star,\quad \forall k\in\mathcal{K}$\;
			$u_{n_{{l}}}^0 \gets u_{n_{{l}}}^\star, \quad \forall n_{{l}}\in\mathcal{N}_{{l}}, \forall {l}\in \mathcal{K} $\;
		}
	}
\end{algorithm}

\section{Unfolded Graph Learning} \label{sec:nn}

Although the networked AirComp can be effectively tackled through conventional optimization techniques as elaborated in the previous section, it requires complicated iterations with convex problem solvers involved to find the solution. As such, the multi-round iterations can be rather time-consuming, which becomes a severe bottleneck when the network scale becomes large. While one of the fundamental motivations of AirComp is efficient data processing, the scalability issue of optimization-based approach may present a significant challenge. In this regard, a well-trained neural network with efficient execution becomes an attractive solution. However, the pure data-drive approach requires large-amount data and large-scale neural networks for training and representation. In this regard, we can refer to the property and structure of the solution from the optimization perspective for specialized neural network architecture design. Also, we can actively address physical interactions within the AirComp network with explicit model to better track the feature of the problem. Therefore, we propose the unfolded deep graph learning framework, where the unfolding leverages the analytical results through optimization technique and the graph neural network models the interactive AirComp network. The unfolding approach integrates domain-knowledge for neural network design, leading to better data efficiency and interpretability as compared with a generic deep model.

\begin{figure*}[t]
	\centering
	\includegraphics[width=0.83\linewidth]{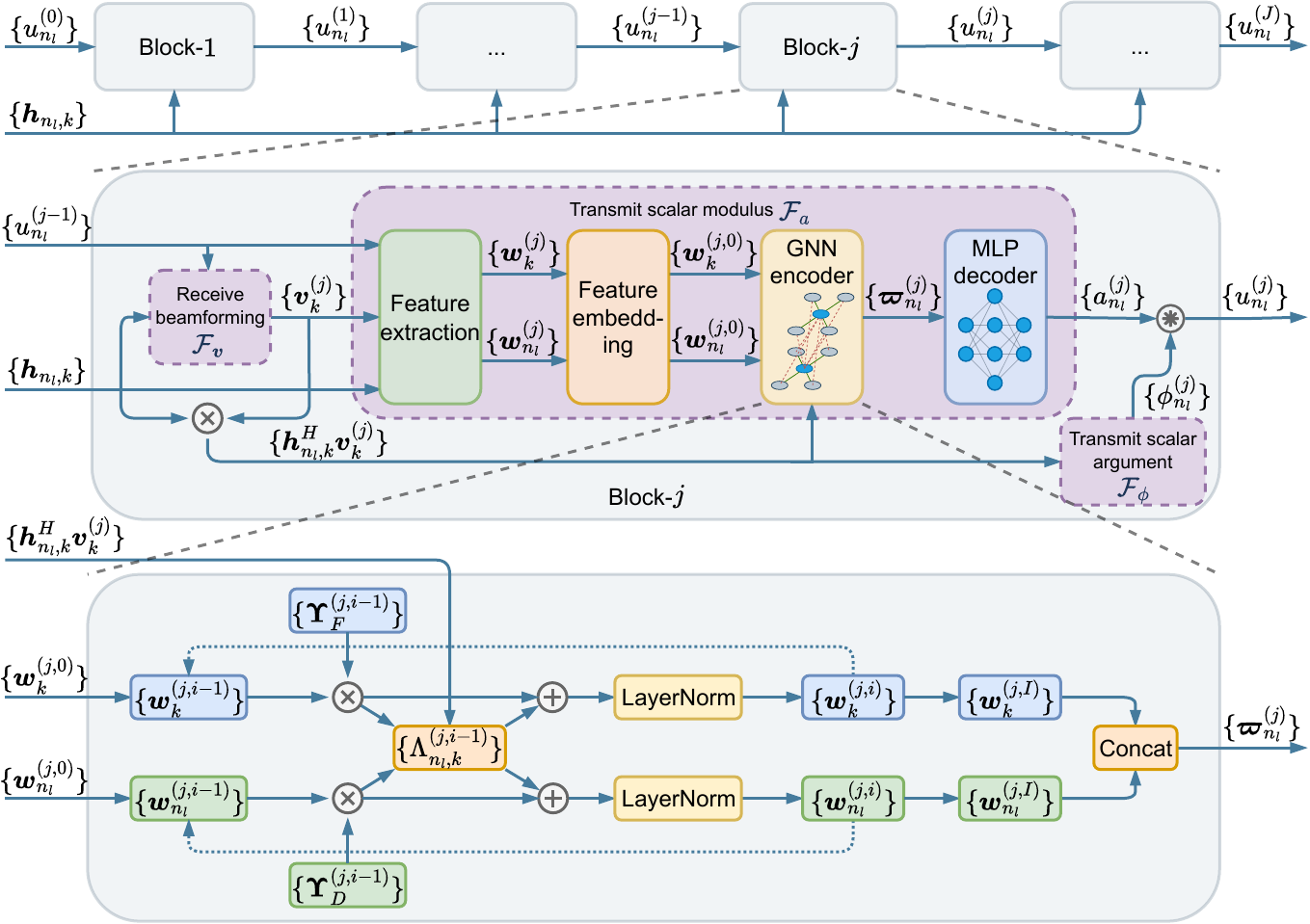}
	\caption{The unfolded deep graph learning architecture.}
	\label{fig:arch}
\end{figure*}

\subsection{Unfolding Architecture}

Based on the previous analysis through optimization, we know that the optimal receive beamformer and transmit scalar can be obtained according to~(\ref{eq:optimal_vk}) and by solving the problem in~(\ref{eq:sp02}), where the former has a closed-form solution while the latter requires an iterative process. In this regard, the receiver beamformer operator in~(\ref{eq:optimal_vk}) can be exploited directly, while the transmit scalar needs additional treatment. Specifically, to tackle the transmit scalars in complex domain, they are rewritten in the form of
\begin{equation} \label{eq:ma}
	u_{n_{{l}}} = a_{n_{{l}}} e^{j\phi_{n_{{l}}}}, \quad \forall n_{{l}}\in\mathcal{N}_{{l}}, \:\: \forall {l}\in \mathcal{K},
\end{equation}
where $ a_{n_{{l}}} $ and $ \phi_k $ are the modulus and argument, respectively. Here, we consider the modulus-argument representation rather than the real and imaginary separations. Although the data dimension are the same, the former is more advantageous since the argument, as can be seen later, can be obtained in a closed form. Moreover, the separation allows real-valued neural network design, which significantly simplifies the implementation and training compared to the case using complex-valued networks. With transmit scalar in~(\ref{eq:ma}), the MSE in~(\ref{eq:msek_ex}) at fusion {center}-$k$ is recast as
\begin{equation}
	\begin{aligned}
		\mathsf{MSE}_k
		=& \sum\limits_{n_k\in\mathcal{N}_k} \left\lvert\vk^H \hnkk \ank e^{j\phi_{n_k}}-1\right\rvert^2\\
		&+ \sum\limits_{l\in\mathcal{K}\backslash\{k\}}\sum\limits_{n_l\in\mathcal{N}_l} \left\lvert\vk^H \hnlk \anl \right\rvert^2
		+\vk^H\vk\sigma_k^2, \forall k\in \mathcal{K}.
	\end{aligned}
	\label{eq:mse2}
\end{equation}
Then, given the monotonic relationship between the AirComp rate and MSE, the rate maximization turns in MSE minimization, which leads to the optimal argument of transmit scalar satisfying
\begin{equation} 
	e^{j\phi_{n_k}}=\frac{\hnkk^H\vk}{\abs{\hnkk^H\vk}}, \quad \forall n_k\in\mathcal{N}_k, \:\: \forall k\in \mathcal{K}.
	\label{eq:arg}
\end{equation}
This optimal phase alignment can be derived by considering the term $|v_{k}^{H}\bm{h}_{n_{k},k}a_{n_{k}}e^{j\phi_{n_{k}}} - 1|^2$ within the MSE expression, for which we align the complex vector $v_{k}^{H}\bm{h}_{n_{k},k}a_{n_{k}}e^{j\phi_{n_{k}}}$ with the positive real direction and obtain the result. It indicates that the transmission phase at each device is determined by the information in the same cluster while being irrelevant with other clusters. Further, by substituting the argument in~(\ref{eq:arg}), the MSE in~(\ref{eq:mse2}) is further derived as
\begin{equation}
	\begin{aligned}
		\mathsf{MSE}_k
		=& \sum\limits_{n_k\in\mathcal{N}_k} \left||\vk^H \hnkk| \ank-1\right|^2\\
		&+ \sum\limits_{l\in\mathcal{K}\backslash\{k\}}\sum\limits_{n_l\in\mathcal{N}_l} \left\lvert\vk^H \hnlk \anl \right\rvert^2
		+\vk^H\vk\sigma_k^2, \forall k\in \mathcal{K},
	\end{aligned}
	\label{eq:mse_a}
\end{equation}
depending on the modulus of the transmit scalar in real domain. Accordingly, we can solve the problem in the form of
\begin{IEEEeqnarray}{cl} \label{eq:sp2}
	\IEEEyesnumber \IEEEyessubnumber*
	\max_{\{a_{n_{{l}}}\}_{n_{{l}}\in \mathcal{N}_{{l}},{l}\in\mathcal{K}}}
	&\quad \sum\limits_{k\in\mathcal{K}} w_k\frac{1}{Q_k+\log_2\Nk}\log_2^+\left(\frac{1}{\mathsf{MSE}_k}\right)
	\\
	\mathrm{s.t.}
	&\quad a_{n_{{l}}}^2 \leq P_{n_{{l}}}\quad \forall n_{{l}}\in\mathcal{N}_{{l}}, \forall {l}\in\mathcal{K},
	\IEEEeqnarraynumspace\IEEEeqnarraynumspace
\end{IEEEeqnarray}
where MSE is given in~(\ref{eq:mse_a}), and the optimization variables are now in real domain.

Technically, we can similarly update the receive beamformer in~(\ref{eq:optimal_vk}), and derive the argument and modulus of the transmission scalars through~(\ref{eq:arg}) and by solving the problem in~(\ref{eq:sp2}), respectively, so as to reach the AirComp solution. However, due to the inefficient nature of iterations, we propose the unfolded learning architecture to approximate the transceiver design. Specifically, we construct a layered neural network to reinterpret the iterative process to update the transmit and receive policies, where the neural network is capable of achieving competent strategy output in an efficient manner.


The proposed unfolded deep graph learning is illustrated in Fig.~\ref{fig:arch}. Overall, denote the collective channel condition of the considered AirComp system as $ \bm{H} = \left[ h_{{n_l}{,k}} \right]_{n_l\in\mathcal{N}_l, k,l \in\mathcal{K}} $, we intend to design a neural network, denoted by parameters $ \bm{\Theta} $, to tackle the input channel condition for an output transceiving strategy, denoted by $ \Phi\left( \bm{H}; \bm{\Theta} \right) $. Then, the achieved weighted-sum {AirComp} rate is obtained as $ R\left( \Phi\left( \bm{H}; \bm{\Theta} \right), \bm{H} \right) $, which approximates the optimum when the neural network is well-trained. Towards this goal, the proposed architecture consists of $ J $ cascaded blocks (also noted as ``layers'' in some unfolded learning researches~\cite{uf-gnn-wmmse,uf-gnn-coord}), where the update in $j$-th block is denoted by $ \Phi^{j} $ with $ \Phi = \Phi^{(J)} \circ \Phi^{(J-1)} \circ \cdots \circ \Phi^{(1)} $. As inspired by the theoretical analysis before, each block corresponds to an iterative updating process among receive beamforming, and argument and modulus of the transmission scalar. Accordingly, the updates in each block consist of three subblocks as\footnote{For the neural network-related discussions, we omit the set notation in the subscript for cleaner representation without {affecting} the rigorousness, e.g., $\{\bm{v}_k\}_{k\in\mathcal{K}}$, ${\{u_{n_k}\}_{n_k\in\mathcal{N}_k}}$, $ \{u_{n_l}\}_{n_l\in\mathcal{N}_l,l\in\mathcal{K}} $, $\{h_{n_l,k}\}_{n_l\in\mathcal{N}_l,l,k\in\mathcal{K}}$ are simplified as $\{\bm{v}_k\}_{k}$, ${\{u_{n_k}\}_{n}}$, $ \{u_{n_l}\}_{n_l} $, $\{h_{n_l,k}\}_{n_l,k}$, respectively.}
\begin{subnumcases}{\label{eq:F}}
	\bm{v}_k^{(j)} = \mathcal{F}_{\bm{v}} \left( \left\{ u_{n_{{l}}}^{(j-1)} \right\}_{n_{{l}}}, {\left\{u_{n_k}^{(j-1)}\right\}_n}; {\bm{H}} \right), \:\: \forall k\in\mathcal{K}, \label{eq:F_v} \\
	e^{j\phi_{n_k}^{(j)}} = \mathcal{F}_{\phi} \left( \bm{v}_k^{({j})};{\bs{h}_{n_k,k}} \right), \quad \forall n_k\in\mathcal{N}_k,\:\:\forall k\in\mathcal{K}, \label{eq:F_phi} \\
	\label{eq:F_a}
	\left\{a_{n_{{l}}}^{(j)}\right\}_{n_{{l}}} = \mathcal{F}_a \left( \left\{ \bm{v}_k^{({j})} \right\}_{k}, {\left\{u_{n_l}^{(j-1)}\right\}_{n_l};} {\bm{H},\Theta^{(j)}}\right),\\
	\label{eq:a_phi}
	{u_{n_l}^{(j)} = a_{n_l}^{(j)}e^{j\phi_{n_l}^{(j)}}, \;\; \forall n_l\in \mathcal{N}_l, \forall l\in \mathcal{K},}
\end{subnumcases}
within the $j$-th blocks, where the connections are built upon their inter-dependencies and shown in Fig.~\ref{fig:arch}. According to~(\ref{eq:F}), the $j$-th block can be denoted as
\begin{equation}
	\left\{ u_{n_l}^{(j)} \right\}_{n_l} = \Phi^{{(j)}} \left( \left\{ u_{n_l}^{(j-1)} \right\}_{n_l}; \bm{H}, \bm{\Theta}^{(j)} \right),
\end{equation}
where $ \bm{\Theta}^{(j)} $ denotes the neural network parameters in this block and the updates of $ \{\bm{v}_k\}_k $ and $ \{\phi_{n_l}\}_{n_l} $ are absorbed. Furthermore, as the updates of $ \mathcal{F}_{\bm{v}} $ and $ \mathcal{F}_{\phi} $ in~(\ref{eq:F_v}) and~(\ref{eq:F_phi}) are conducted according to~(\ref{eq:optimal_vk}) and~(\ref{eq:arg}), respectively, the closed-form results enables convenient computation. For the purpose of optimizing transmit scalar modulus simultaneously, we exploit a GNN with inherent message passing mechanism $\mathcal{F}_{a}$ to model the mutual interactions, as elaborated below.

\subsection{GNN for Transmission Design}

The proposed GNN in each block intends to approximate the optimal modulus of the transmission scalars in current iteration, i.e., implementing the subblock $ \mathcal{F}_a $ in~(\ref{eq:F_a}). In consistence with the physical AirComp network, the GNN is comprised of two types of nodes, for the devices and fusion centers, respectively, leading to a heterogeneous graph. The heterogeneous structure is deliberately chosen rather than a homogeneous one that naturally fits the different types of nodes, allowing more specialized and potentially more effective learning of the AirComp process. Here we do not explicitly consider the edges in a typical graph to achieve an agile model. The nodes are established with particularly designed features, which are then fed into the message-passing GNN encoder for aggregation and propagation among the nodes for updated representations. Finally, a multilayer perceptron (MLP) decoder is used to convert the neural network output for the transmission strategy.

\subsubsection{Feature Extraction and Embedding}

For the proposed GNN to solve the modulus of the transmission scalars, the intended output is in the real domain, while the input, including the channel conditions, previous-round transmission and receive strategies, is in complex domain. To avoid the complexity of tackling complex variables, we first extract the features of interest and embed them as the GNN encoder input. Specifically for block-$j$, from the input transmission scalars $ \{u_{n_l}^{(j-1)}\}_{n_l} $, we first obtain the receive beamformer $ \{\bm{v}_{k}^{(j)}\}_{k} $ through $ \mathcal{F}_{\bm{v}} $. Then, together with channel condition, we extract the features of the device node-$n_l$ and fusion center node-$k$ as
\begin{equation} \label{eq:wnk}
	\begin{aligned}
		\bm{w}_{n_l}^{(j)} = & \Bigg\{\abs{u_{n_l}^{(j-1)}}, \abs{ \bm{h}_{n_l,l}^H \bm{v}_l^{(j)} },\sum\limits_{k\in\mathcal{K}\backslash\{l\}}\abs{ \bm{h}_{n_l,k}^H \bm{v}_k^{(j)} },\\
		&\quad\abs{ \bm{h}_{n_l,l}^H \bm{v}_l^{(j)} u_{n_l}^{(j-1)}}, \sum\limits_{k\in\mathcal{K}\backslash\{l\}}\abs{ \bm{h}_{n_l,k}^H \bm{v}_k^{(j)} u_{n_l}^{(j-1)} } \Bigg\}, \\
		& \qquad\qquad\qquad \qquad\qquad\qquad \forall n_l\in\mathcal{N}_l,\:\:\forall l\in\mathcal{K},
	\end{aligned}
\end{equation}
and
\begin{equation} \label{eq:wk}
	\begin{aligned}
		&\bm{w}^{(j)}_k = \Bigg\{\mathsf{MSE}^{(j-1)}_k,\sum_{n_k\in\mathcal{N}_k}\abs{\hnkk^H \vk^{(j)}},\\
		&\qquad\qquad\:\: \sum_{l\in\mathcal{K}\backslash\{k\}} \sum_{n_l\in\mathcal{N}_l} \abs{\hnlk^H \vk^{(j)}}, \sum_{n_k\in\mathcal{N}_k}\abs{\hnkk^H \vk^{(j)} u_{n_k}^{(j-1)} }, \\
		&\qquad\qquad\:\: \sum_{l\in\mathcal{K}\backslash\{k\}} \sum_{n_l\in\mathcal{N}_l} \abs{\hnlk^H \vk^{(j)}u_{n_l}^{(j-1)}}
		\Bigg\}, \quad \forall k\in\mathcal{K},
	\end{aligned}
\end{equation}
respectively, in the $j$-th block. Different from most existing works that extract the real and imaginary parts of the complex input to facilitate real-domain neural networks, the extracted features in~(\ref{eq:wnk}) and~(\ref{eq:wk}) aggregate the complex input into more concise representations, where each node is associated with a five-tuple feature. Moreover, the extracted features are of clear physical interpretations that approximately correspond to the useful and interfering channel conditions. The underlying implication is that, the AirComp rate calculation depends on the combined effects from the signal and interference and thus we do not need to bother probing into each single channel.

Then, the extracted features are embedded as the initialization for the GNN encoder. In this regard, the nodes of the same type share a neural network for embedding, denoted by $ f_{D}^{(0)} $ and $ f_{F}^{(0)} $ for the device nodes and fusion center nodes with network parameters $ \bm{\Omega}_{D}^{(j)} $ and $ \bm{\Omega}_F^{(j)} $, respectively, in the $j$-th block. As such, the embedding operation for device node-$n_l$ and fusion center node-$k$ are specified as
\begin{equation} \label{eq:agg_nl}
	\bm{w}_{n_l}^{(j,0)} = f_{D}^{(0)} \left( \bm{w}_{n_l}^{(j)}; \bm{\Omega}_{D}^{(j)} \right) = \mathsf{tanh} \left( \bm{\Omega}_{D}^{(j)} \bm{w}_{n_l}^{(j)} \right),
\end{equation}
and
\begin{equation} \label{eq:agg_k}
	\bm{w}^{(j, 0)}_k = f_F^{(0)} \left( \bm{w}^{(j)}_k; \bm{\Omega}_F^{(j)} \right) = \mathsf{tanh} \left( \bm{\Omega}_F^{(j)} \bm{w}^{(j)}_k \right),
\end{equation}
respectively, where the superscript-0 indicates the initialization layer for the GNN encoder, and $ \mathsf{tanh} $ is exploited as the activation function.


\subsubsection{GNN Encoder}

The GNN encoder tackles the embedded features of nodes by exploiting the message passing mechanism such that the features can be sufficiently fused to represent the network characteristics, which further facilitates the strategic transmission output. To this end, the encoder consists of $ I $ layers of message passing, as shown in Fig.~\ref{fig:arch}. Specifically, the message passing consists of message generation, aggregation and combination as cascaded linear layers, where the operation in the $i$-th layer can be represented as
\begin{equation}
\begin{aligned}
	\bm{w}_{n_l}^{(j,i)} &= f_{n_l}^{(i)} \bigg( \bm{w}_{n_l}^{(j,i-1)}, \left\{\bm{w}_{k}^{(j,i-1)}\right\}_k;. \\
	& \qquad\qquad\qquad\quad \bm{\Upsilon}_{D}^{(j,i)} , \left\{ {\Lambda}_{n_l,k}^{(j)} \right\}_{n_l,k} \bigg) \\
	&= \mathsf{LN} \bigg( \bm{\Upsilon}_{D}^{(j,i)} \bm{w}_{n_l}^{(j,i-1)}\\
	& \qquad\qquad\qquad + \sum\limits_{k\in\mathcal{K}} {\Lambda}_{n_l,k}^{(j)} \bm{\Upsilon}_{F}^{(j,i)} \bm{w}_{k}^{(j,i-1)} \bigg),
\end{aligned}
\end{equation}
and
\begin{equation}
\begin{aligned}
	\bm{w}^{(j, i)}_k &=  f_k^{(i)} \bigg( \bm{w}^{(j, i-1)}_k, \left\{ \bm{w}_{n_l}^{(j,{i})} \right\}_{n_l};\\
	 & \qquad\qquad\qquad\qquad\bm{\Upsilon}_{F}^{(j, i)}, \bm{\Upsilon}_{D}^{(j, i)},  \left\{ {\Lambda}_{n_l,k}^{(j)} \right\}_{n_l,k}  \bigg) \\
	&= \mathsf{LN} \bigg( \bm{\Upsilon}_{F}^{(j, i)} \bm{w}^{(j, i-1)}_k \\
	& \qquad\qquad\quad +  \sum\limits_{l\in\mathcal{K}} \sum\limits_{n_l\in\mathcal{N}_l} {\Lambda}_{n_l,k}^{(j)} \bm{\Upsilon}_{D}^{(j, i)} \bm{w}_{n_l}^{(j,{i})} \bigg),
\end{aligned}
\end{equation}
where $ f_{n_l}^{(i)} $ and $  f_k^{(i)} $ denote message passing with respect to the input node features of the devices and fusion centers, respectively, $ \bm{\Upsilon}_{D}^{(j, i)} $ and $  \bm{\Upsilon}_{F}^{(j, i)} $ are the shared neural network parameters for message generation at the devices and fusion centers, respectively, in the $i$-th encoder layer in the $j$-th block. Then, for the message aggregation, each node aggregates the features of nodes of the other type through weighed sum-pooling, i.e., device node aggregates the features of fusion center nodes and vice versa, where the weight coefficients $ \left\{ {\Lambda}_{n_l,k}^{(j)} \right\}_{n_l,k} $ are defined as
\begin{equation}
	\Lambda_{n_l,k}^{(j)}=\left\{
	\begin{aligned}
		&\abs{\hnlk^H\vk^{(j)}}, &\quad l=k,\\
		&-\abs{\hnlk^H\vk^{(j)}},&\quad \text{otherwise}.
	\end{aligned}\right.
\end{equation}
Finally, the nodes combine their previous feature and the obtained aggregated message through a linear layer with a LayerNorm function denoted by $\mathsf{LN}(\cdot)$. The layer normalization operates across the feature dimension for each node's representation independently, assisting to stabilize the activation output and improve training dynamics within the GNN encoders.

For the proposed message passing mechanism in the GNN encoder, there are two-fold specialized designs. First, by introducing the coefficients $ \left\{ {\Lambda}_{n_l,k}^{(j)} \right\}_{n_l,k} $, the aggregation results are positively enhanced with respect to the intended signal, while negatively affected against the interference. Second, we define the message passing that the devices node exchange information with all fusion centers but without the other devices, and vice versa. This is different from most of the existing GNN-based frameworks where the message aggregation is conducted over all the neighboring nodes, and thus alleviates the communication overhead for the GNN training and execution.

After the $I$-layer message passing, the representation of all nodes are well updated and we concatenate the features of each device node with the features of its associated fusion center as the encoder output, given as
\begin{equation}
	\bm{\varpi}^{(j)}_{n_l} = \mathsf{Concat}\left( \bm{w}_{n_l}^{(j,I)}, \bm{w}^{(j, I)}_l \right).
\end{equation}
Therefore, the overall GNN encoding network operation can be represented by
\begin{equation}
	\bm{\varpi}^{(j)}_{n_l} = \Phi_{\text{encd}}^{(j)} \left( \left\{ \bm{w}^{(j, 0)}_k\right\}_k, \left\{ \bm{w}_{n_l}^{(j,0)} \right\}_{n_l}; \bm{H}, \bm{\Theta}^{(j)}_{\text{encd}} \right),
\end{equation}
where $ \bm{\Theta}^{(j)}_{\text{encd}} $ collects the GNN parameters in the $j$-th block, and
\begin{equation}
\begin{aligned}
	\Phi_{\text{encd}}^{(j)} = &\mathsf{Concat}\circ \bigg(  \left\{ f_k^{(I)} \circ f_k^{(I-1)} \circ \cdots \circ f_k^{(1)} \right\}_k,\\
	& \qquad\qquad \:\left\{ f_{n_l}^{(I)} \circ f_{n_l}^{(I-1)} \circ \cdots \circ f_{n_l}^{(1)} \right\}_{n_l} \bigg).
\end{aligned}
\end{equation}

\subsubsection{MLP Decoder}
With GNN-encoded output features for each device node, we adopt a MLP decoder to further interpret the intended modulus of the transmission scalar at each device. Particularly, the decoder interpretation can be represented as
\begin{equation}
	a_{n_l}^{{(j)}} = \sqrt{P_{n_l}} \Phi_{\text{decd}}^{(j)} \left( \bm{\varpi}^{(j)}_{n_l} ; \bm{\Theta}^{(j)}_{\text{encd}} \right),
\end{equation}
where $ \Phi_{\text{decd}}^{(j)} $ denotes the decoding operation with $ \bm{\Theta}^{(j)}_{\text{encd}} $ being the neural network parameters. Moreover, we adopt the scaled exponential linear units ($\mathsf{SELU}$) as the activation for the input and hidden layers and $\mathsf{Sigmoid}$ as the output activation function, which is further scaled to the interval $ \left[ 0, \sqrt{P_{n_l}} \right] $ corresponding to the transmission scalar modulus.

As the final operation in the $j$-th block, the modulus of the transmission scalar obtained from $\mathcal{F}_a$-subblock and the argument from $\mathcal{F}_{\phi}$-subblock constitute the transmission scalar, denoted as $ \left\{ u_{n_l}^{(j)} \right\}_{n_l} $, according to~(\ref{eq:a_phi}). The updated transmission scalar vector is then fed into the next block within the unfolded learning architecture until the last one.

As a further note, our proposed framework synergistically combines algorithm unfolding with GNNs, where the deep unfolding mimics the optimization-inspired iterations and GNN captures the complex and graph-structured interference dependencies. We expect the proposed unfolded graph learning framework to learn an efficient mapping from channel conditions to the transceiver design to achieve optimized networked AirComp performance.

\subsection{Model Training}

For the unfolded graph learning architecture introduced above, we need to train the network parameters such that the weighted-sum AirComp rate can be maximized. Accordingly, we define the loss function as
\begin{equation}
	\mathsf{L} (\bm{\Theta}) = -\mathbb{E}_{\bm{H}\sim\mathcal{H}} \left\{ R\left( \Phi\left( \bm{H}; \bm{\Theta} \right), \bm{H} \right) \right\},
\end{equation}
where $ \mathcal{H} $ denotes the conforming distribution of the collected channel conditions. We adopt the unsupervised learning that, given the samples of channel conditions $\bm{H}$, the unfolded learning network generates the transceiving strategy $ \Phi\left( \bm{H}; \bm{\Theta} \right) $ to be trained for the highest average weighted-sum AirComp rate for random deployments, i.e., the minimization of the loss function. The stochastic gradient descent approach is employed for neural network training, where the real-domain calculation significantly facilitates the training process.

Moreover, to improve the training efficiency and performance, we adopt the \textbf{progressive learning} and \textbf{parameter sharing} policies. Due to the non-differential positivizing operation of $ (\cdot)^+ $ in AirComp rate, the gradient calculation for the loss function can possibly be problematic. Specifically, when the MSE is higher than 1, the function output is 0, leading to zero gradients being backpropagated for the cluster contribution. To tackle this issue, we adopt the progressive learning technique, where the training is divided into two stages. In the first stage, the positivizing operation is ignored, using a differentiable loss to ensure informative gradients and guide the network towards a good parameter region. Accordingly, the neural network is further enhanced in the second stage by explicitly considering the original positivizing operation as a fine-tuning process towards the true objective. Also, the learning rate is scheduled with an exponential decay, which allows larger steps during the initial-stage training and more refined ones when approaching the optimum, in order to achieve improved convergence. Moreover, for the proposed block-wise learning architecture, the parameter sharing policy is employed such that the first-half blocks share the same parameters, and so do the rest blocks. The underlying assumption is that, the operations within the early-stage iterations have inherent similarities, so do the latter iterations. In this regard, the number of model parameters is significantly reduced and the training can be finished in a more memory- and computation-efficient manner.

\section{Simulation Results} \label{sec:sim}

\subsection{Simulation Settings}

The evaluated network is defined within a circular area with a radius of $2000~\text{m}$. There are $5$ clusters randomly deployed in the network, the associated $5$ devices are randomly located within the area between the radius of $100~\text{m}$ and $1000~\text{m}$. The cluster center is deployed with a fusion center, which is equipped with $8$~antennas to receive the AirComp signals. The channel between the devices and fusion centers follow Rician fading, where the path loss exponent is $2.2$ with $-30~\text{dB}$ attenuation at the reference distance. The Rician factor is $5$ with line-of-sight component and Rayleigh non-line-of-sight component. The noise power is $-90~\text{dBm}$, and maximum transmission power each wireless device is $1~\text{W}$. The weights for AirComp rate is set to $1$ for all clusters. The parameters above are used as the default network setting for training unless otherwise noted.

For the neural network, we consider $6$ cascaded blocks, where the GNN encoder in each block has $2$ layers for massage passing. The parameters in GNN encoder, $ \{ \bm{\Omega}_{\{\cdot\}}^{(\cdot)} \} $ and $ \{ \bm{\Upsilon}_{\{\cdot\}}^{(\cdot,\cdot)} \} $ are of dimensions $ 32\times5 $ and $ 32\times32 $, respectively, and the MLP decoder parameters are of dimensions $[64, 1000, 500, 32, 1]$. The are $3000$ training epochs and the learning rate is initialized as $ 5\times10^{-5}$ with an exponent decaying coefficient of $0.9$.

The benchmark schemes includes:
\begin{itemize}
	\item \textbf{UDGL}: The proposed unfolded deep graph learning. Thanks to the inherent GNN structure, the trained model under default setting can be directly generalized to new scenarios.
	\item \textbf{UDGL(Re)}: The proposed approach with further refinement. For new scenarios, the neural network trained under default setting is exploited as a pre-trained model, and then further fine tuning is conducted with new scenario parameters to reach a refined model.
	\item \textbf{UDGL(Tr)}: The proposed unfolded deep graph learning with truncated unfolding architecture that consists of only two blocks.
	\item \textbf{UMLP}: The unfolded learning scheme through pure MLP structure without GNN. Note the MLP-based model is basically not generalizable and thus requires re-training when encountering new scenarios.
	\item \textbf{AO}: The proposed alternating optimization algorithm.
	\item \textbf{FPT}: The devices transmit with their maximum power.
	\item \textbf{APT}: The adaptive transmission scheme that the devices with worse channel conditions are of higher power, scaled according to the poorest channel, given as
	\begin{equation}
		\unk= \sqrt{P_{n_k}}\frac{\min\limits_{m_k\in\mathcal{N}_k}\{\abs{\bm{h}_{m_k,k}^H \bm{v}_k}\}\hnkk^H\vk}{\abs{\hnkk^H\vk}^2}, \quad \forall n_k,
	\end{equation}
	where $ \frac{\min_{m_k\in\mathcal{N}_k}\{\abs{\bm{h}_{m_k,k}^H \bm{v}_k}\}}{\abs{\hnkk^H\vk}} $ is the scaling factor inducing the device with the worst channel to transmit with its maximum power.
\end{itemize}

\subsection{Convergence and Model Size}

\begin{figure}[t]
	\centering
	\includegraphics[width=0.9\linewidth]{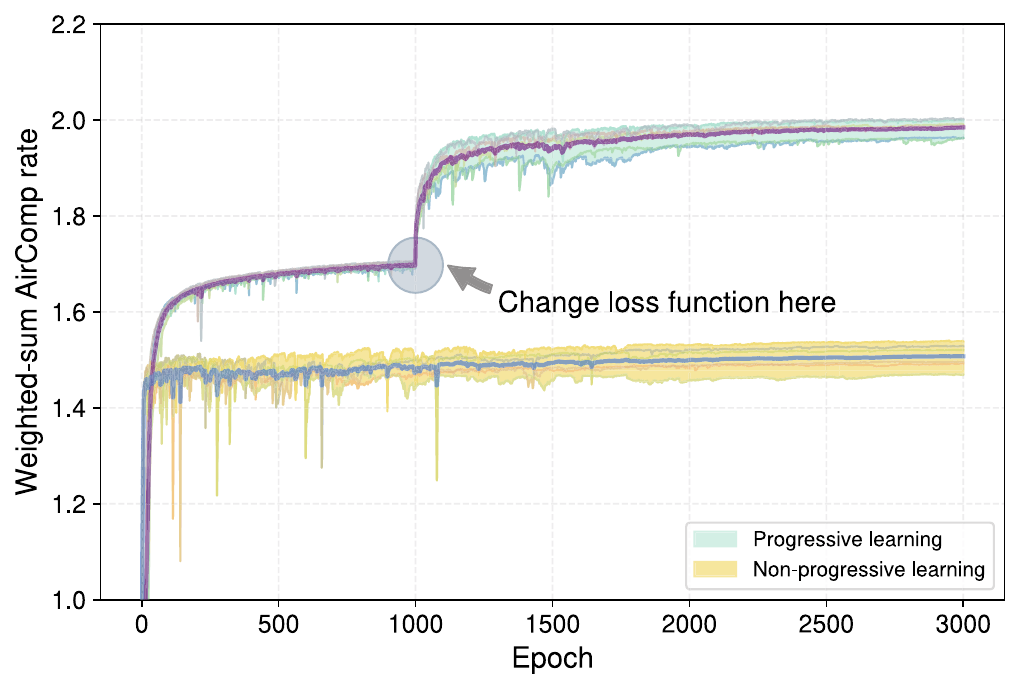}
	\caption{Achieved AirComp rate while training for the cases with and without progressive learning.}
	\label{fig:loss}
\end{figure}

In Fig.~\ref{fig:loss}, we show the convergence of model training under the proposed unfolded learning with trials of different seeds. The cases with and without progressive learning are considered where the former corresponds to our proposal while the latter keeps using the non-differential AirComp rate while training. As we can see, our proposal with progressive learning in the first stage adopts the differential loss and can be well trained with accurate gradients, and the performance is even improved in the second stage when considering the positivizing operation. The progressive learning here can also be regarded as fine-tuning, where the first-stage results are employed as a pre-trained model that is further refined in the second stage with a changed loss function. In contrast, the non-progressive learning directly use the non-differential AirComp rate as loss, and thus the potential inaccurate gradients affect the training process and downgrade the performance.

In~Table~\ref{tab:size}, we consider the different settings of the unfolding neural network size along with the achieved performance, where the training and inference time under the smallest-scale network are used as the reference. Evidently, the network setting achieving the highest performance is exploited as the default, with reasonable cost for training and execution. Moreover, it needs to note that when the model size becomes even larger, the performance may not necessarily gets further improved but possibly downgraded. This reveals the potential over-fitting issue and training difficulties incurred with large-size neural networks.

\begin{table}[t]\small
	\centering
	\caption{Performance under different model sizes}
	\begin{tabularx}{\linewidth}{>{\centering\arraybackslash}c
			>{\centering\arraybackslash}c
			>{\centering\arraybackslash}X
			>{\centering\arraybackslash}X
			>{\centering\arraybackslash}X}
	\toprule
	$J$                & $I$ & AirComp rate & Train time & Inference time \\ \hline
	\multirow{3}{*}{$4$} & $1$ & $1.87$ & $100~\%$    & $100~\%$        \\
	& $2$ & $1.91$ & $106~\%$    & $106~\%$        \\
	& $3$ & $1.83$ & $112~\%$    & $113~\%$        \\
	\hline
	\multirow{3}{*}{$6$} & $1$ & $1.96$ & $151~\%$    & $147~\%$        \\
	& $2$ & $\bf{1.98}$ & ${160~\%}$    & ${156~\%}$        \\
	& $3$ & $1.86$ & $171~\%$    & $168~\%$        \\
	\hline
	\multirow{3}{*}{$8$} & $1$ & $1.96$ & $203~\%$    & $197~\%$        \\ 
	& $2$ & $1.95$ & $215~\%$    & $207~\%$        \\
	& $3$ & $1.92$ & $228~\%$    & $223~\%$        \\ \bottomrule
\end{tabularx} \label{tab:size}
\end{table}

\subsection{Performance Comparison}

We evaluate the performance of different schemes under different network settings. Here, we emphasize that, among the learning-based scheme, our proposals employ the GNN structure and thus can be directly generalized to different scenarios \textbf{without retraining}. Particularly, we train the deep unfolded learning neural network with the default simulation settings presented in Sec.~\ref{tab:size}. Then, the results for the cases with different numbers of clusters, devices, and antennas are obtained by directly reusing the trained neural network under default settings without specially re-training a dedicated neural network. In contrast, the learning scheme with only the MLP is not capable for generalization, and the results under different settings require \textbf{newly trained} neural networks.

\begin{figure}[t]
	\centering
	\includegraphics[width=0.9\linewidth]{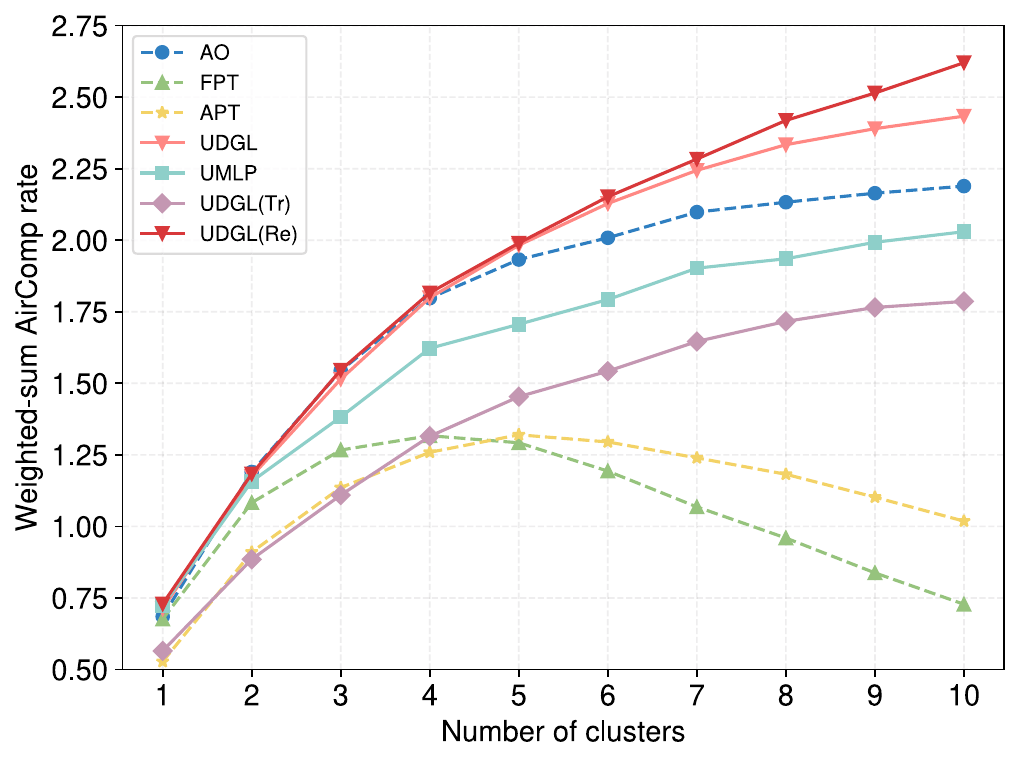}
	\caption{Aircomp rate vs. number of clusters.}
	\label{fig:clu}
\end{figure}

\begin{figure}[t]
	\centering
	\includegraphics[width=0.9\linewidth]{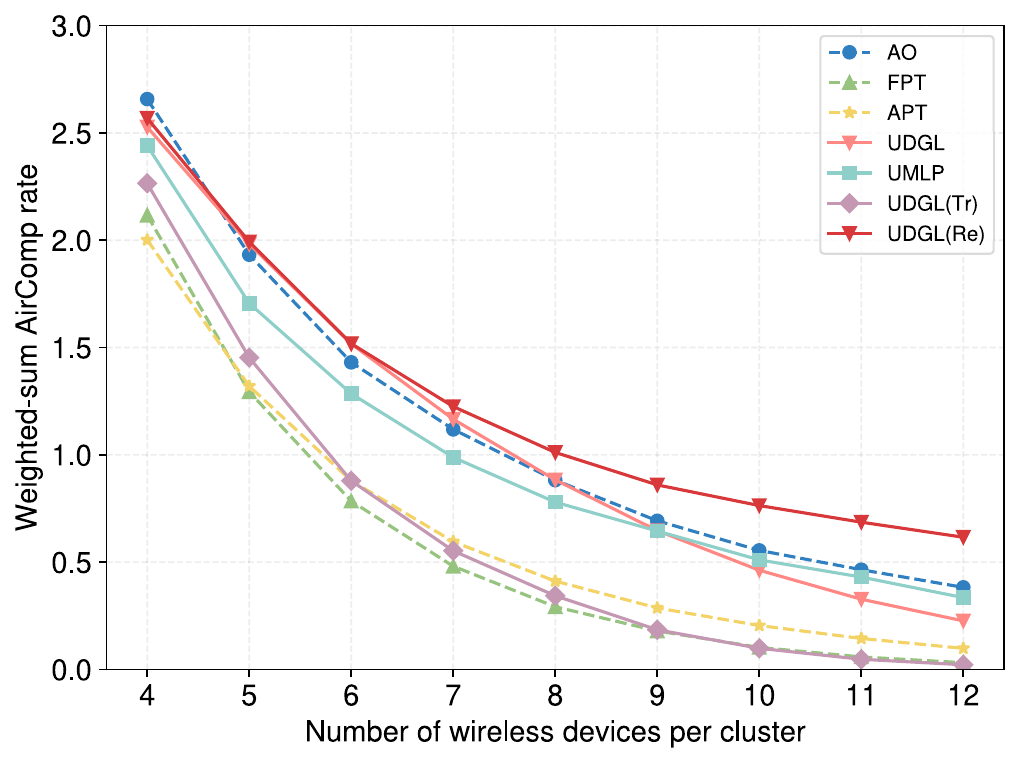}
	\caption{AirComp rate vs. number of devices in each cluster.}
	\label{fig:node}
\end{figure}

\begin{figure}[t]
	\centering
	\includegraphics[width=0.9\linewidth]{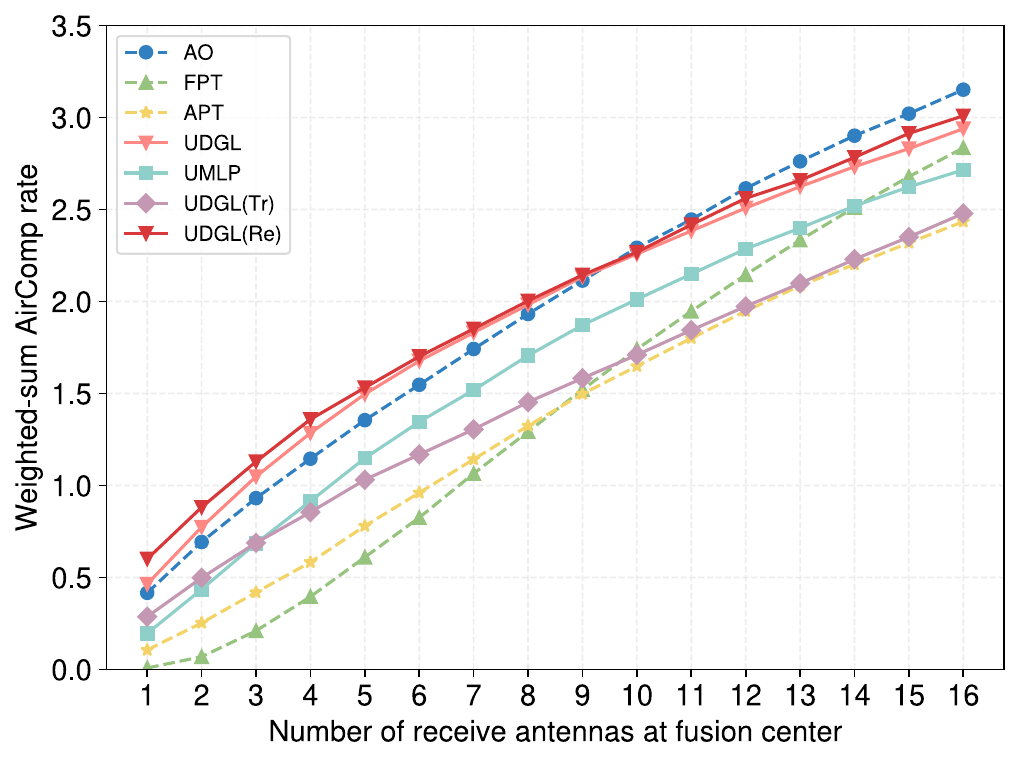}
	\caption{AirComp rate vs. number of receive {antennas} at each fusion center.}
	\label{fig:ant}
\end{figure}

In Fig.~\ref{fig:clu}, we show the achieved AirComp rate under different schemes against the number of clusters. Generally, the AirComp rate increases but gradually tends saturated with an increasing number of clusters, except for the cases with adaptive transmission and full-power transmission that induce decreased rate when there coexist more than 4 clusters. Particularly, the proposed learning scheme achieves better performance as compared with the conventional alternating optimization scheme for the case with 5 clusters where the neural network is trained. Moreover, for the cases with more clusters where the results under our proposal are directly generalized from the trained neural network, the advantage of our proposal compared with the optimization approach becomes more significant. The reason is that the optimization approach is only capable of reaching the stationary points as suboptima, possibly deviating far from the optimum. This phenomenon becomes even more severe with more clusters where the heavy interference induces large number of stationary points. In contrast, the learning-based scheme is capable to extract the featured representation of the AirComp network with improved performance. In this regard, proper design of the neural network structure becomes prominent, as evidenced by our proposal. Moreover, we can see that the fine-tune of trained network for changed scenario further helps enhance the performance. Conversely, for the MLP-based proposal, which fails specialized design to track the node interactions, achieved limited performance despite its capability for universal approximation. Also, for the truncated unfolding structure with two blocks, the small-size model only has limited representation capability and thus induces poor performance. Additionally, regarding the cases with adaptive power and full power, both are significantly dominated by other schemes. The full-power transmission performs relatively well for the cases with {fewer number of clusters} and thus slighter interference, and the inverse holds for the cases with adaptive power transmissions.

In Fig.~\ref{fig:node}, we evaluate the performance with changing number of devices per cluster. Generally, as the number of devices in each cluster increases, the difficulty of signal alignment rises, and interference intensifies, resulting in a sharp decline in the AirComp rate. Similarly to the results in Fig.~\ref{fig:clu}, our proposal with fine-tuning is capable of achieving the better performance as compared with the others, including the optimization scheme. When without fine-tuning, the proposed unfolded learning may not generalize sufficiently well and thus can be dominated by the optimization scheme. Furthermore, the AirComp rate is sensitive to changes in the number of devices. When the number of devices becomes large, the network features differ significantly from those under the training parameter settings, resulting in a generalized rate that is lower than that obtained by retraining the MLP-based method using the parameters from inference.

In Fig.~\ref{fig:ant}, the performance evaluation is conducted by considering the number of receive antennas at the fusion centers. Generally, with an increasing number of antennas, the receiving AirComp capability of the fusion centers becomes strengthened and thus the AirComp rate is increased. Specifically, the proposed unfolded learning outperforms other schemes at the training point and generalizes well for the cases with relatively small number of antennas. However, when the fusion centers are equipped sufficient number of antennas, e.g., more than 10, the optimization-based approach is capable for more superior performance as compared with our proposal. The more antennas there are, the greater the freedom of receive beamforming, allowing for more effective interference suppression, and $\frac{1}{\mathsf{MSE}}$ typically falls within the logarithmic region of the $\log^+$ function. The AirComp rate loss due to relaxing $\log^+$ to a $\log$ function in the AO method is negligible. Moreover, as AO iterations increase, it becomes harder for neural networks with limited blocks to approximate this process.

\begin{figure}[t]
	\centering
	\includegraphics[width=0.9\linewidth]{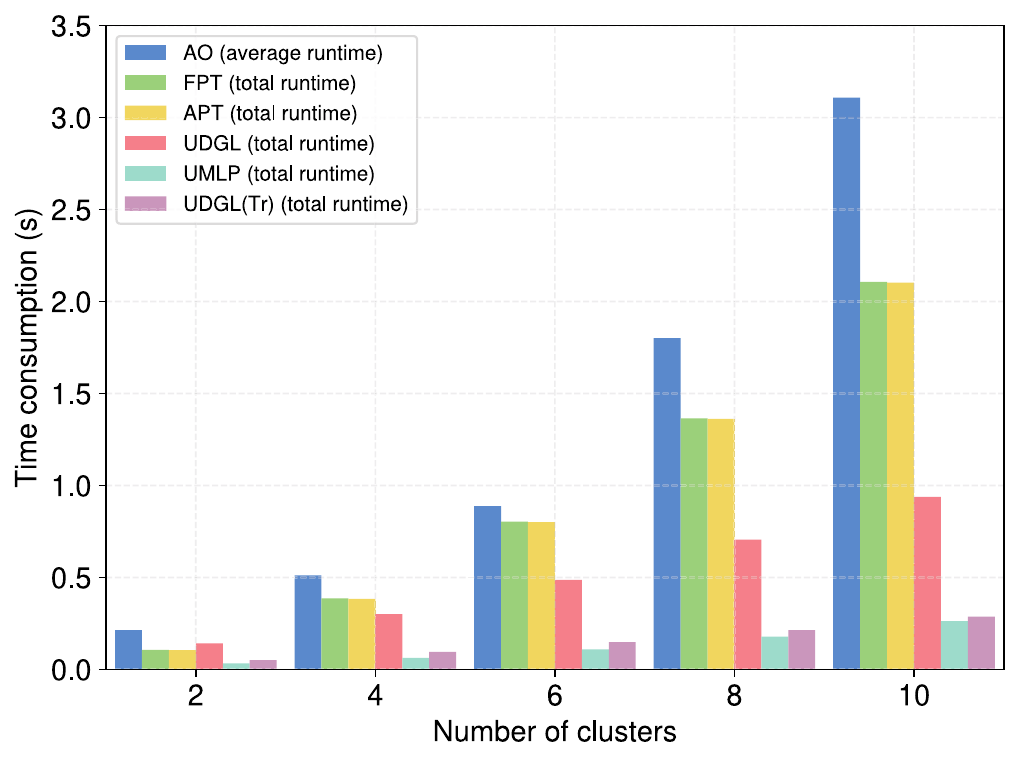}
	\caption{Time consumption for algorithm execution vs. number of clusters.}
	\label{fig:time}
\end{figure}

In Fig.~\ref{fig:time}, we demonstrate the time consumption for algorithm execution under different approaches. Generally, the alternating optimization requires large numbers of iterations, the adaptive-power and full-power schemes are nearly in closed-form, and the learning-based approaches can be promptly executed. Accordingly, we present the average runtime of the alternating optimization scheme over 5000 random deployments, as well as the total runtime of the rest schemes under the same settings. The results are obvious that the neural networks almost have instantaneous output, and far more efficient than iterative optimization. Overall, from Figs.~\ref{fig:clu},~\ref{fig:node},~\ref{fig:ant} and ~\ref{fig:time}, we can see that the proposed unfolded learning has evident performance superiority, and the advantage can be further enhanced with fine tuning. Moreover, due to the inherent GNN structure adopted, the proposed model has effective generalization capability to cover untrained cases. The generalization capability inherently stems from the permutation equivalence property of the proposed GNN structure, enabling the reuse of a trained network to new scenarios without tr-trained specialized model. Also, the generalization demonstrates particular advantages for the cases with heavy interference, where the optimization approach may plunge into the suboptima and thus dominated by the proposed learning scheme.

\begin{figure}[t]
	\centering
	\includegraphics[width=0.9\linewidth]{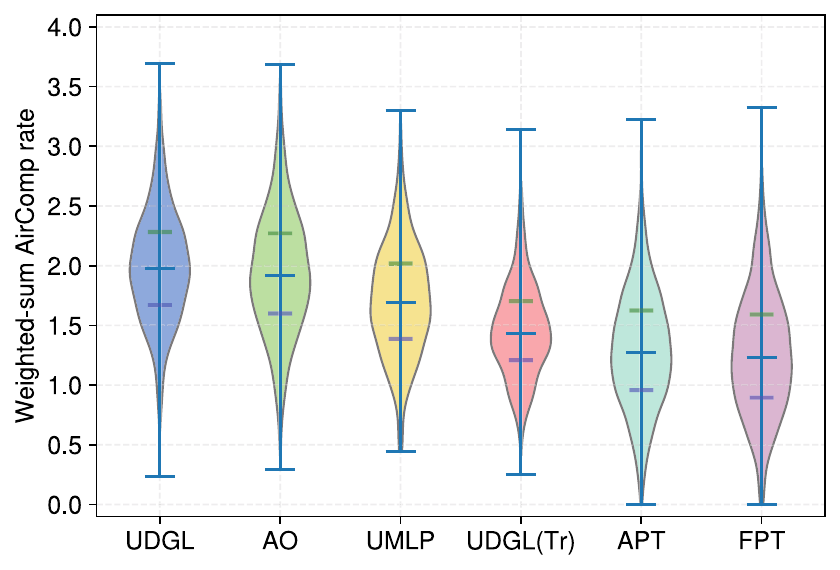}
	\caption{Distribution of AirComp rate with 5 clusters (through training).}
	\label{fig:violin5}
\end{figure}

\begin{figure}[t]
	\centering
	\includegraphics[width=0.9\linewidth]{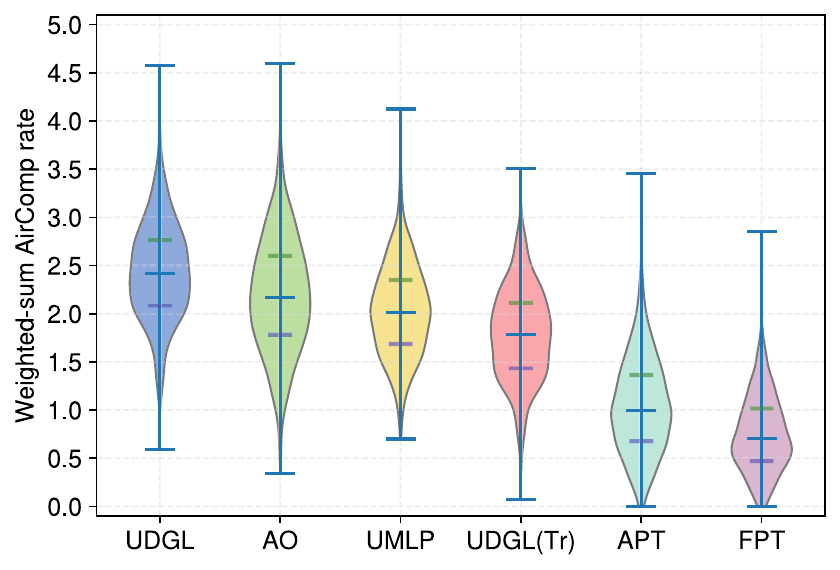}
	\caption{Distribution of AirComp rate with 10 clusters (through generalization).}
	\label{fig:violin10}
\end{figure}

Besides the network-wide overall performance, we also demonstrate some micro-scope performance below. In Fig.~\ref{fig:violin5}, we show the distribution of achieved AirComp rate under different schemes for the case with 5 clusters, where the neural network are trained under this setting. Generally, the 5-cluster case corresponds to the scenario with relatively slight/medium interference. In consistence with the results presented before, the proposed unfolded learning and alternating optimization achieve comparable results. The achieved distribution under two schemes also resembles each other, whereas the learning scheme obtains slightly higher median. Furthermore, we evaluate the case with 10 clusters, as shown in Fig.~\ref{fig:violin10}, where the unfolding network trained under 5-cluster case is directly exploited without retraining. Note the case with 10 clusters is for the scenarios with heavy interference. As we can see, although obtained through generalization without dedicated re-training, our proposed model achieves the leading performance. The distribution of achieved rate is superior under our proposal and so are the median and quartiles, as compared with the optimization-based scheme, let alone the rest approaches. Therefore, the unfolding model learns the inherent ability for interference management, and the ability keeps preserved during generalization. When further inspecting the performance under MLP-based learning and truncated unfolded learning, we can see that the problem-specific neural network structure design and proper network scale build the foundation towards an effective learning model.

\begin{figure}[t]
	\centering
	\includegraphics[width=\linewidth]{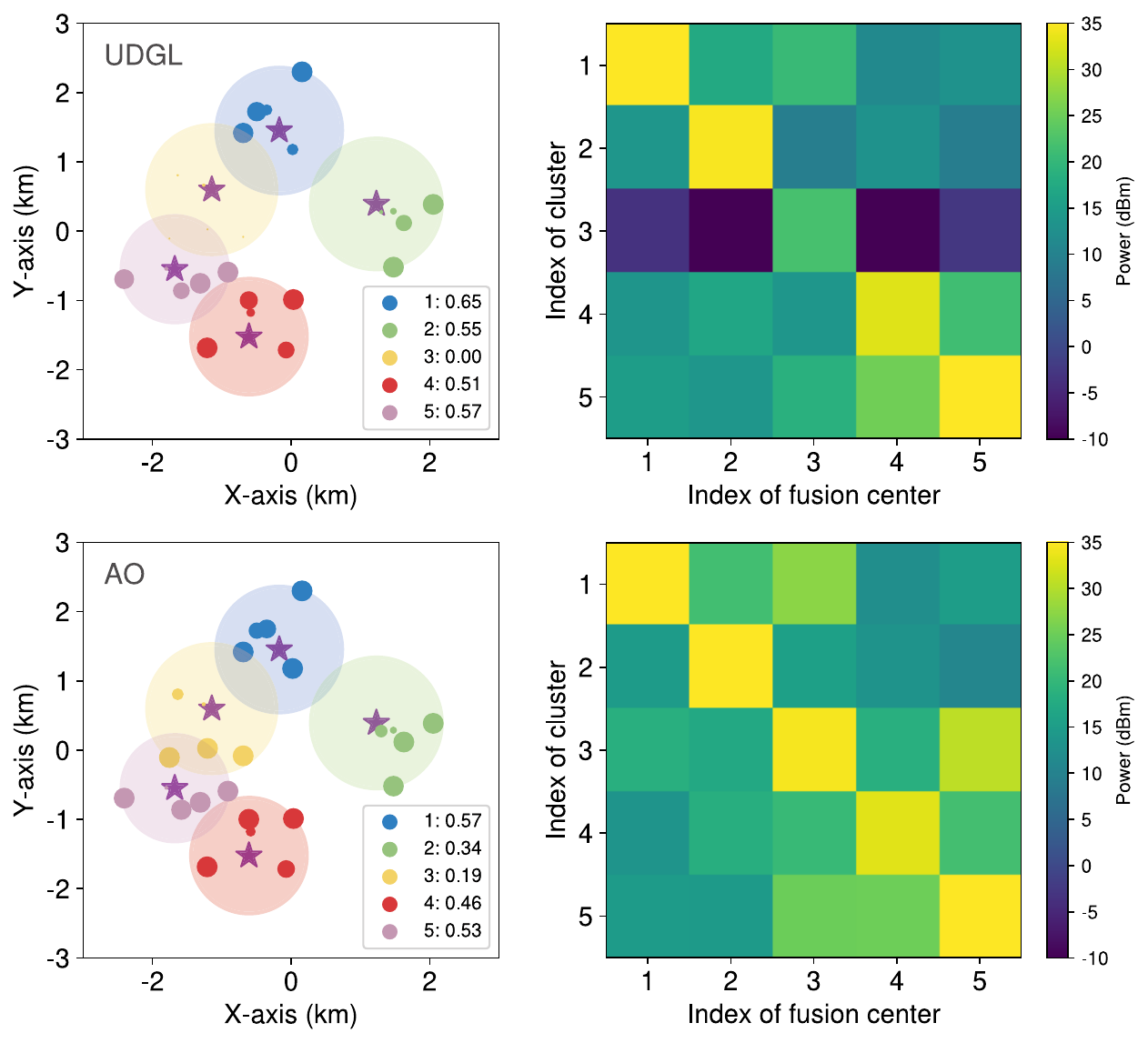}
	\caption{Test demonstration of transmit power and mutual interference with 5 clusters.}
	\label{fig:trial}
\end{figure}

Furthermore, we consider a specific simulation test for the case with 5 clusters, for which the power and interference strength are shown in Fig.~\ref{fig:trial}. Specifically, the upper subfigures correspond to results from the proposed unfolded learning, and lower subfigures for the optimization-based approach. Also, the left subfigures demonstrate the topology with 5 clusters where the bubble size indicates the transmit power strength, and the right subfigures show the power strength of the intra-cluster signal (diagonal elements) and inter-cluster interference (non-diagonal elements). We can see that in this test, compared with the results through optimization, our proposal suppresses the transmissions in cluster-3, and thus the transmissions in other clusters are of relatively higher power. Accordingly, the power heat map also indicates that under our proposal, the lower transmit power of cluster-3 alleviates the interference posed upon other clusters. We show the achieve AirComp rate in each cluster as specified by the legend in the lower right corner of the left subfigures. Therefore, our proposal induces relatively lower AirComp rate in cluster-3, but the AirComp rate in other clusters are improved, which further leads to higher overall AirComp rate compared to the result from the optimization approach.

\subsection{Transfer Learning}

The preceding results have revealed the sound capability for generalization of the proposed unfolded learning approach, we further evaluate its transfer learning capability to different scenarios of physical networks. Particularly, we consider the different setting in terms of network area radius (where the distance among the nodes in the network is scaled with the same ratio), maximum transmit power, and noise power. Since the evaluations for these changes do not necessarily alter the neural network architecture, we evaluate the performance under our proposal and the MLP-based neural network, and exploit the results through optimization as the reference. Here, the transfer learning refer to that, the neural networks trained under the default setting are used as the pre-trained models, for both unfolded graph learning and MLP-based learning. Then, for the new scenario settings, a small number of samples are used to fine tune the pre-trained model such that the new scenarios can be sufficiently fitted. Also, note that the idea of transfer learning is generally not applicable to the optimization-based approach, where any change in setting requires the same procedure of re-calculation.

The results in Tables~\ref{tab:rad},~\ref{tab:pwr},~\ref{tab:noise} details the transfer learning performance with respect to network area radius, maximum transmit power, and noise power, respectively. From the figures in the tables, we can see that compared with the MLP-based learning, the proposed unfolded learning achieves better results, for both direct application of trained model and transferred model in new scenarios. Meanwhile, when the system is allowed with higher freedom for interference management, i.e., larger areas, higher transmit power, or lower noise power, the optimization approach may exceed our proposal that is directly applied new scenarios. When the contrary holds and the AirComp network incurs higher interference, the results through our proposed learning outperform those through optimization, which is in consistence with the results indicated before. Therefore, transfer learning based on the pre-trained model empowers better adaptation to the new scenarios with significant convenience, and thus achieved improved AirComp rate, for which a well-established basic model allows greater potential in the changing environments.

\begin{table}[!t]\small
	\centering
	\caption{Transfer learning for different network area radii}
	\begin{tabularx}{\linewidth}{>{\centering\arraybackslash}c
			>{\centering\arraybackslash}c
			>{\centering\arraybackslash}c
			>{\centering\arraybackslash}X}
		\toprule
		Area radius      & Scheme & Trained/Applied   & Transferred \\ \hline
		\multirow{3}{*}{\shortstack{$2.0$ km\\(Training point)}} & UDGL   & $\bf{1.98}$ & --                 \\
		& UMLP   & ${1.70}$ & --                 \\
		& AO    & ${1.93}$ & --                    \\ \hline
		\multirow{3}{*}{$1.5$ km} & UDGL   & $\bf{1.67}$ & $\bf{1.69}$                 \\
		& UMLP   & $1.44$ & $1.58$                 \\
		& AO    & $1.56$ & N.A.                    \\ \hline
		\multirow{3}{*}{$2.5$ km} & UDGL   & ${2.22}$ & $\bf{2.24}$                 \\
		& UMLP   & $1.92$ & $2.03$                 \\
		& AO    & $\bf{2.23}$ & N.A.                    \\
		\bottomrule
	\end{tabularx} \label{tab:rad}
\end{table}

\begin{table}[!t]\small
	\centering
	\caption{Transfer learning for different maximum allowed power}
	\begin{tabularx}{\linewidth}{>{\centering\arraybackslash}X
		>{\centering\arraybackslash}c
		>{\centering\arraybackslash}c
		>{\centering\arraybackslash}c}
		\toprule
		Max transmit power       & Scheme & Trained/Applied   & Transferred \\ \hline
		\multirow{3}{*}{\shortstack{$1$ W\\(Training point)}}   & UDGL   & $\textbf{1.98}$ & --                \\
		& UMLP   & ${1.70}$ & --                 \\
		& AO     & ${1.93}$ & --                    \\ \hline
		\multirow{3}{*}{$0.5$ W} & UDGL   & $\bf{1.49}$ & $\bf{1.51}$                 \\
		& UMLP   & $1.24$ & $1.35$                 \\
		& AO     & $1.37$ & N.A.                    \\ \hline
		\multirow{3}{*}{$2$ W}   & UDGL   & $2.41$ & $\bf{2.44}$                 \\
		& UMLP   & $2.08$ & $2.22$                 \\
		& AO     & $\bf{2.46}$ & N.A.                    \\
		\bottomrule
	\end{tabularx} \label{tab:pwr}
\end{table}

\begin{table}[!t]\small
	\centering
	\caption{Transfer learning for different noise power}
	\begin{tabularx}{\linewidth}{>{\centering\arraybackslash}c
		>{\centering\arraybackslash}c
		>{\centering\arraybackslash}c
		>{\centering\arraybackslash}X}
		\toprule
		Noise power                & Scheme & Trained/Applied   & Transferred \\ \hline
		\multirow{3}{*}{\shortstack{$-90$ dBm\\(Training point)}} & UDGL   & $\bf{1.98}$ & --                 \\
		& UMLP   & ${1.70}$ & --                 \\
		& AO    & ${1.93}$ & --                    \\ \hline
		\multirow{3}{*}{$-100$ dBm} & UDGL   & $3.11$ & $\bf{3.24}$                 \\
		& UMLP   & $2.59$ & $2.98$                 \\
		& AO    & $\bf{3.51}$ & N.A.                    \\ \hline
		\multirow{3}{*}{$-110$ dBm} & UDGL   & $3.49$ & $\bf{3.80}$                 \\
		& UMLP   & $2.72$ & $3.61$                 \\
		& AO    & $\bf{4.56}$ & N.A.                    \\
		\bottomrule
	\end{tabularx} \label{tab:noise}
\end{table}

\section{Conclusion} \label{sec:con}

In this paper, we study the networked AirComp issue with an unfolded deep graph learning framework, where the unfolding architecture with an inherent GNN is used to track the mutual influence among multi-cluster AirComp. The results demonstrate that the proposed learning scheme achieves superior performance as compared with the conventional approaches. Moreover, the GNN-based structure enables strong generalization capability to different network scenarios, where the learned capability of interference management empowers superior performance to enhance the networked AirComp.

\bibliographystyle{IEEEtran}
\bibliography{main}

\end{document}